\documentclass[sigconf]{acmart}
\makeatletter
\def\@ACM@checkaffil{
    \if@ACM@instpresent\else
    \ClassWarningNoLine{\@classname}{No institution present for an affiliation}%
    \fi
    \if@ACM@citypresent\else
    \ClassWarningNoLine{\@classname}{No city present for an affiliation}%
    \fi
    \if@ACM@countrypresent\else
        \ClassWarningNoLine{\@classname}{No country present for an affiliation}%
    \fi
}
\makeatother

\AtBeginDocument{%
  \providecommand\BibTeX{{%
    \normalfont B\kern-0.5em{\scshape i\kern-0.25em b}\kern-0.8em\TeX}}}

\usepackage{xcolor}
\definecolor{s5}{RGB}{255, 234, 221}
\definecolor{5t10}{RGB}{252, 174, 174}
\definecolor{10t20}{RGB}{255, 137, 137}
\definecolor{l20}{RGB}{255, 102, 102}

\definecolor{best1}{RGB}{122,189,129}
\definecolor{best2}{RGB}{149,198,132}
\definecolor{best3}{RGB}{208,235,140}
\definecolor{best4}{RGB}{239,230,144}
\definecolor{worst1}{RGB}{250,225,143}
\definecolor{worst2}{RGB}{241,179,130}
\definecolor{worst3}{RGB}{237,159,124}
\definecolor{worst4}{RGB}{231,114,111}

\usepackage{multirow}
\usepackage{subfigure}
\usepackage[flushleft]{threeparttable}
\usepackage{url}
\usepackage{amsmath,amsfonts}

\newcommand\doubleplus{+\kern-1ex+}

\copyrightyear{2023} 
\acmYear{2023} 
\setcopyright{rightsretained} 
\acmConference[SEC '23]{The Eighth ACM/IEEE Symposium on Edge Computing}{December 6--9, 2023}{Wilmington, DE, USA}
\acmBooktitle{The Eighth ACM/IEEE Symposium on Edge Computing (SEC '23), December 6--9, 2023, Wilmington, DE, USA}
\acmDOI{10.1145/3583740.3628442}
\acmISBN{979-8-4007-0123-8/23/12}

\begin{document}

\title[Unveiling Energy Efficiency in Deep Learning]{Unveiling Energy Efficiency in Deep Learning: Measurement, Prediction, and Scoring across Edge Devices}

\author{Xiaolong Tu\textsuperscript{1} Anik Mallik\textsuperscript{2} Dawei Chen\textsuperscript{3} Kyungtae Han\textsuperscript{3} Onur Altintas\textsuperscript{3} Haoxin Wang\textsuperscript{1} Jiang Xie\textsuperscript{2}}

\affiliation{
  \institution{\textsuperscript{1}Georgia State University \textsuperscript{2}The University of North Carolina at Charlotte \textsuperscript{3}Toyota InfoTech Labs}
}

\email{{xtu1, haoxinwang}@gsu.edu, {amallik, linda.xie}@uncc.edu, {dawei.chen1, kt.han, onur.altintas}@toyota.com}

\renewcommand{\shortauthors}{Xiaolong Tu et al.}

\begin{abstract}
Today, deep learning optimization is primarily driven by research focused on achieving high inference accuracy and reducing latency.
However, the energy efficiency aspect is often overlooked, possibly due to a lack of sustainability mindset in the field and the absence of a holistic energy dataset.
In this paper, we conduct a threefold study, including energy measurement, prediction, and efficiency scoring, with an objective to foster transparency in power and energy consumption within deep learning across various edge devices.
Firstly, we present a detailed, first-of-its-kind measurement study that uncovers the energy consumption characteristics of on-device deep learning.
This study results in the creation of three extensive energy datasets for edge devices, covering a wide range of kernels, state-of-the-art DNN models, and popular AI applications.
Secondly, we design and implement the first kernel-level energy predictors for edge devices based on our kernel-level energy dataset.
Evaluation results demonstrate the ability of our predictors to provide consistent and accurate energy estimations on unseen DNN models.
Lastly, we introduce two scoring metrics, PCS and IECS, developed to convert complex power and energy consumption data of an edge device into an easily understandable manner for edge device end-users. 
We hope our work can help shift the mindset of both end-users and the research community towards sustainability in edge computing, a principle that drives our research. Find data, code, and more up-to-date information at https://amai-gsu.github.io/DeepEn2023.
\end{abstract}



\maketitle

\section{Introduction}
\label{sc:Introduction}
Recently, there has been heavy investment in implementing various AI applications on mobile and edge devices, for instance, (1) \textit{vision-based} AI applications, such as image classification \cite{howard2017mobilenets,krizhevsky2017imagenet,szegedy2016rethinking}, face recognition \cite{li2015convolutional,schroff2015facenet}, object detection and tracking \cite{redmon2016you,huang2017speed,7001050}, image super-resolution \cite{dong2015image,ledig2017photo,timofte2018ntire}, segmentation \cite{chen2018encoder}, pose estimation \cite{papandreou2018personlab}, and gesture recognition \cite{ordonez2016deep}; (2) \textit{natural language processing} (NLP) based applications, such as smart reply \cite{serban2017deep}, question answering \cite{rajpurkar2016squad}, language translation \cite{bahdanau2014neural,sutskever2014sequence}, and sentiment analysis \cite{severyn2015twitter,socher2013recursive}; and (3) \textit{voice-based} applications, such as virtual-assistant \cite{kepuska2018next}, speech recognition \cite{chiu2018state}, and sound classification \cite{zhang2015robust}. 

Despite the remarkable advances in edge device capabilities such as functionality, computation power, and storage capacity, the limited energy capacity has been the major bottleneck in promoting advanced edge AI applications.
On one hand, edge AI applications, particularly those that involve intensive computing resources such as deep learning algorithms, tend to consume a significant amount of energy \cite{9787708,wang2020user}.
On the other hand, mobile and edge devices are typically powered solely by embedded batteries, so their energy capacity is significantly constrained by form factor requirements, safety considerations, manufacturing costs, and concerns on the environmental impact of the battery technology used.
As a result, heavy battery usage of an application often results in low ratings or subpar user experience. 
A survey \cite{surveybattery} finds that about $55\%$ of users surveyed would give a negative review to a mobile application that consumes a lot of battery, indicating that energy consumption is a crucial aspect of the user experience that cannot be overlooked.
These observations raise intuitive questions: \textit{
How can we identify the energy bottlenecks and optimize the energy efficiency of on-device deep learning for diverse edge devices?
What are the primary factors that have a large impact on the energy consumption of deep neural network (DNN) executions, the core of on-device deep learning?
Where is the energy spent inside a DNN execution?} 
Answering these questions, however, is challenging, due to the lack of holistic understanding of the intricacies of power and energy consumption in DNN executions on edge devices.
First and foremost, \textit{we cannot optimize what cannot be measured.} The energy efficiency of an edge device is more than its AI hardware capability in isolation. Instead, it is coupled with the on-device deep learning software stack, whose net performance is shrouded beneath the DNN models and end-to-end processing pipeline of diverse edge AI applications. 
Second, \textit{we cannot optimize what is under-appreciated or neglected in the design.} Most existing research and development in deep learning primarily aim to reduce inference latency and enhance accuracy, often neglecting to consider the impact on energy efficiency. 
As a result, it becomes crucial to strike a balance between improving energy efficiency and enhancing performance in on-device deep learning for modern edge devices.

In this paper, we study the problem of accurate energy measurement, prediction, and understandable scoring of on-device deep learning, and make three concrete contributions towards enabling \textit{transparency of power and energy consumption inside on-device deep learning across diverse edge devices}.

First, we conduct the first detailed measurement study to accurately quantify the energy consumed by on-device deep learning across diverse modern edge devices. Our measurement study covers three dimensions, including the power and energy consumption of kernels
, state-of-the-art (SOTA) DNN models, and widely-used edge AI applications. Our measurements reveal multiple key observations, which remain consistent across eight different measured edge devices. 
Overall, we measure and collect fine-grained power traces and accurate energy consumption data for (1) $16$ types of kernels with $1,847$ unique configurations, (2) nine SOTA DNN models with $50$ variants each, and (3) six widely-used edge AI applications on eight commercial edge devices executed with mobile CPU and GPU. These measurements result in creation of three large-scale power and energy datasets, including kernel-, model-, and application-level datasets for on-device deep learning on edge devices.

Second, based on our kernel-level energy dataset and the observations gained in the measurement study, we design and implement kernel-level energy predictors on both mobile CPU and GPU. To the best of our knowledge, this is the first energy predictor for on-device deep learning on commercial edge devices (e.g., modern smartphones), which can provide consistently accurate energy estimation on unseen DNN models. This offers an effective approach to extend our measurements and observations derived from a limited DNN model space to new DNN models, which enhances the extensibility of our measurement study.

Lastly, beyond valuing research that aims at improving the energy efficiency of on-device deep learning, it is crucial that our measurement study are accessible to a wide audience, such as end-users with non-technical backgrounds. For instance, presenting an energy efficiency score, ranging from $0$ to $100$, should be more straightforward and easier to understand than telling end-users that their device will consume $120.090$ mJ per inference to run MobileNetv1 with CPUs.
To this end, we develop two scoring metrics: \textit{power consumption score (PCS)} and \textit{inference energy consumption score (IECS)}. These two scoring metrics help to distill the power and energy efficiency of an edge device in an intuitive and understandable way.
We present a complete scoring results for eight edge devices benchmarked by leveraging our application-level dataset.

\section{Background and Challenges}
\label{sc:challenges}
\subsection{Background}
DNN models are the core of on-device deep learning and consume a major portion of both computational and energy resources on mobile and edge devices.
A DNN model consists of a sequence of primitive operations, such as convolution2D (\texttt{conv}), depthwise convolution2D (\texttt{dwconv}), activations, pooling, and fully-connected (\texttt{fc}), which are organized into layers, allowing the network to learn complex patterns from input data. To enhance the computational efficiency of the DNN inference (i.e., to reduce inference latency and avoid redundant memory access), kernel fusion (or operator fusion) is a key optimization and has been incorporated in SOTA DNN execution frameworks, such as TVM \cite{chen2018tvm}, TFLite \cite{abadi2016tensorflow}, and MNN \cite{jiang2020mnn}.
For instance, three individual operations, \texttt{conv}, batch normalization (\texttt{bn}), and rectified linear unit (\texttt{relu}) can be fused into one composite operation, \texttt{conv$\doubleplus$bn$\doubleplus$relu}\footnote{In this paper, $\doubleplus$ represents kernel fusion.}, to achieve inference acceleration on edge devices. This means the entire sequence can be processed as a single step, which reduces memory access (since intermediate results don't need to be written to and read from memory) and kernel launch overhead. Hence, given its crucial role in runtime optimization, a kernel is typically considered as the fundamental unit for scheduling and execution in deep learning frameworks, particularly on edge devices \cite{zhang2021nn}.

\subsection{Challenges}
\textbf{C1: Accuracy.}
In order to optimize the energy efficiency of DNN executions on resource-constrained edge devices, it is crucial to gain a deep understanding of the energy consumption characteristics associated with various DNN models across different edge hardware platforms, such as mobile CPUs and GPUs. Consequently, the importance of conducting accurate measurement studies on real devices is becoming increasingly paramount. However, measuring accurate energy consumption on a real edge device is non-trivial. The challenges arise from two main observations: (1) existing energy profiling methods for edge devices, which rely on built-in current sensors, cannot capture power consumption at high time granularity (i.e., less than 100 ms); and (2) the growing level of integration in the electronic circuits of edge devices presents challenges when attempting to connect them with an external power monitor.

\begin{table}[t]
\centering
\caption{MobileNetv1 energy consumption.}
\vspace{-0.1in}
\small
\begin{tabular}{ccccc}
\toprule
\multirow{2}{*}{} & \multicolumn{2}{c}{CPU} & \multicolumn{2}{c}{GPU} \\ 
\cline{2-5} 
                  & Energy  &  Error  &  Energy  &   Error      \\ 
\hline 
Built-in & 132.420mJ  &  10.3\%  &  19.254mJ  &   30.64\%      \\   
\hline
Ground-truth  & 120.090mJ  &  -  &  27.760mJ  &  -  \\   
\bottomrule
\end{tabular}
\label{tb:model}
\end{table}

\begin{table}[t]
\centering
\caption{MobileNetv1 individual kernel energy consumption.}
\vspace{-0.1in}
\small
\begin{tabular}{cccc}
\toprule
\multirow{2}{*}{Kernels} & \multicolumn{3}{c}{CPU} \\ 
\cline{2-4} 
                  & Built-in (mJ)  &  Ground-truth (mJ)  &  Error  \\ 
\hline 
conv$\doubleplus$relu                &3.914 & 3.984 & \colorbox{s5}{1.76\%}   \\   
dwconv$\doubleplus$relu               &5.578 & 4.814 & \colorbox{10t20}{15.8\%}   \\   
conv$\doubleplus$relu                &8.020 & 7.739 & \colorbox{s5}{3.62\%}   \\   
dwconv$\doubleplus$relu                &8.682 & 8.193 & \colorbox{5t10}{5.96\%}   \\   
conv$\doubleplus$relu                &2.649 & 2.422 & \colorbox{5t10}{9.36\%}   \\   
dwconv$\doubleplus$relu                &5.211 & 4.428 & \colorbox{10t20}{17.6\%}   \\   
conv$\doubleplus$relu                &1.225 & 0.930 & \colorbox{l20}{31.8\%}   \\   
dwconv$\doubleplus$relu                &1.541 & 1.285 & \colorbox{l20}{20.0\%}   \\   
conv$\doubleplus$relu                &2.030 & 1.643 & \colorbox{l20}{23.5\%}   \\   
dwconv$\doubleplus$relu                &7.824 & 6.549 & \colorbox{10t20}{19.5\%}   \\   
conv$\doubleplus$relu                &3.450 & 2.933 & \colorbox{10t20}{17.6\%}   \\   
dwconv$\doubleplus$relu                &0.174 & 0.149 & \colorbox{10t20}{16.8\%}   \\   
conv$\doubleplus$relu                &1.179 & 0.972 & \colorbox{l20}{21.3\%}   \\   
dwconv$\doubleplus$relu                &2.879 & 2.448 & \colorbox{10t20}{17.6\%}   \\   
conv$\doubleplus$relu                &12.394& 11.324& \colorbox{5t10}{9.45\%}   \\   
dwconv$\doubleplus$relu                &0.524 & 0.466 & \colorbox{10t20}{12.4\%}   \\   
conv$\doubleplus$relu                &14.112& 12.976& \colorbox{5t10}{8.76\%}   \\   
dwconv$\doubleplus$relu                &0.906 &  0.771& \colorbox{10t20}{17.5\%}   \\   
conv$\doubleplus$relu                &12.065& 11.095& \colorbox{5t10}{8.74\%}   \\   
dwconv$\doubleplus$relu                &1.108 & 0.944 & \colorbox{10t20}{17.3\%}   \\   
conv$\doubleplus$relu                &14.446& 13.327& \colorbox{5t10}{8.39\%}   \\   
dwconv$\doubleplus$relu                &0.409 & 0.367 & \colorbox{10t20}{11.5\%}   \\   
conv$\doubleplus$relu                &12.240& 11.357& \colorbox{5t10}{7.77\%}   \\   
dwconv$\doubleplus$relu                &0.299 & 0.267 & \colorbox{10t20}{11.7\%}   \\   
conv$\doubleplus$relu                &4.349 & 4.019 & \colorbox{5t10}{8.20\%}   \\   
dwconv$\doubleplus$relu                &0.110 & 0.093 & \colorbox{10t20}{17.6\%}   \\   
conv$\doubleplus$relu                &4.353 & 3.902 & \colorbox{10t20}{11.6\%}   \\   
global-pool                &0.071 & 0.062 & \colorbox{10t20}{14.4\%}   \\   
fully connected                &0.664 & 0.636 & \colorbox{s5}{4.42\%}   \\   
\bottomrule
\end{tabular}
\begin{tabular}{llll}
\textcolor{s5}{$\blacksquare$} error $\leq 5\%$ & \textcolor{5t10}{$\blacksquare$} $5\% <$ error $\leq 10\%$ &\\
\textcolor{10t20}{$\blacksquare$} $10\% <$ error $\leq 20\%$ & \textcolor{l20}{$\blacksquare$} error $>20\%$
\end{tabular}
\label{tb:kernel}
\end{table}

\begin{figure*}[t]
\centerline{\includegraphics[width=1.0\textwidth]{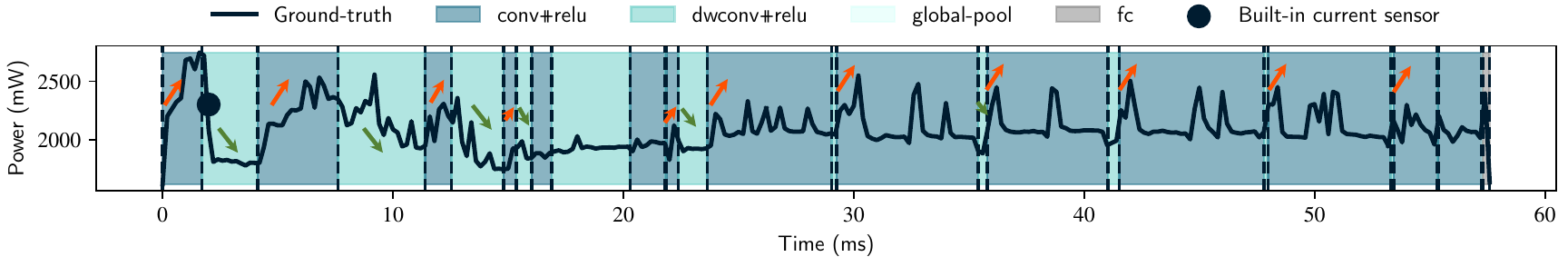}}
\vspace{-0.1in}
\caption{Comparison of time-granularity between the device's built-in current sensor and external power monitor.}
\label{fig:p1_challenge}
\end{figure*}

\begin{figure}[t]
\centering
\subfigure[Samsung Galaxy S5 with a snap-type battery connector (2014)]
{\includegraphics[width=0.23\textwidth]{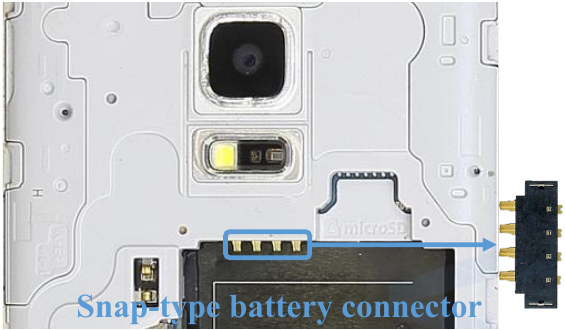}\label{fig:S5}}
\subfigure[Samsung Galaxy S20 with an FPC battery connector (2020)]
{\includegraphics[width=0.23\textwidth]{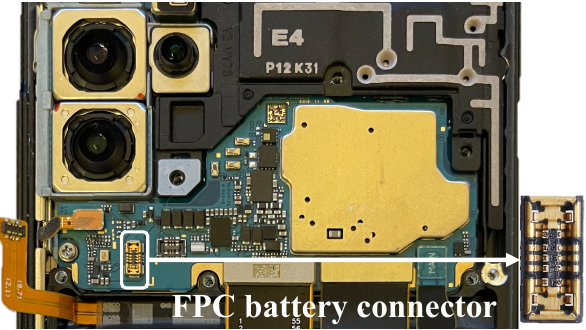}\label{fig:S20}}
\vspace{-0.15in}
\caption{Comparison between an older snap-type battery connector and a modern FPC connector.}
\label{fig:batteryconnector}   
\end{figure}

First, most SOTA DNN models can achieve inference latencies of 10 to 200 ms when executed on mobile CPUs. These latencies can be significantly reduced to a range of 1 to 50 ms when executed on mobile GPUs \cite{zhang2021nn}. On the other hand, a DNN model usually consists of tens or hundreds of kernels that run sequentially on the edge device \cite{cai2018proxylessnas, dai2019chamnet, zhang2021nn}, each potentially having an execution time of less than a millisecond. 
Therefore, to accurately capture the instantaneous power variations within a DNN inference, which includes the precise power consumption of individual kernels, an ideal power sampling rate should be less than $1$ ms. However, we have observed that existing edge devices, such as smartphones, typically have built-in current sensors (e.g., fuel gauge) with a time-granularity of approximately $100$ ms to $1$ second. This restricts the sampling rate at which the sensors can measure the power drawn by the device to $1-10$ times per second. This indicates that the existing built-in current sensors cannot fully capture the fine-grained, kernel-level power variations within a DNN inference on the edge device, resulting in inaccurate measurements.

We have conducted a measurement study on a real device, Huawei P40 Lite, to investigate the extent of this discrepancy compared to the ground-truth power and energy consumption\footnote{In this paper, the ground-truth power and energy consumption is measured by connecting the real device to the Monsoon power monitor \cite{Monsoon}.}. As shown in Tables \ref{tb:model} and \ref{tb:kernel}, measurements dependent on the device's built-in current sensor produce large errors in both the overall DNN model ($10.3\% - 30.64\%$) and individual kernel ($1.76\% - 31.8\%$) energy consumption\footnote{Energy consumption is calculated by multiplying the measured power consumption by the model/kernel inference latency. To ensure the energy consumption errors are primarily caused by the power measurement inaccuracy, we use the ground-truth latency for calculating the energy consumption in built-in current sensor measurements.}. Moreover, as we show in Section \ref{sc:pred}, using the energy dataset created by a built-in current sensor to train an energy predictor results in consistently poor estimation accuracy. In addition, Fig. \ref{fig:p1_challenge} demonstrates the built-in sensor fails to capture the characteristics of power variations among kernels within a DNN inference.
For instance, there is usually a sudden power rise at the start of \texttt{conv} executions, and a sudden power drop in most \texttt{dwconvs}.

\textit{Consequently, these observations indicate that existing energy profiling solutions for edge devices that heavily rely on the built-in current sensor may fail to offer accurate power measurements for DNN executions (e.g., power profiler \cite{jindal2021experience}, reading virtual file \texttt{current\_now} from \texttt{/sys/class/power\_supply/battery/} \cite{apicharttrisorn2019frugal}, and reading battery level drops from \texttt{ACTION\_BATTERY\_CHANGED} \cite{ran2018deepdecision,chen2015smartphone}).}

Second, one of the common methods to measure accurate and fine-grained power consumption for mobile and edge devices in the research community is to connect the device to an external power monitor with a high sampling rate \cite{pathak2012energy,mallik2023epam,hindle2014greenminer,9274509,mcintosh2019can,wang2019energy}. However, we find that connecting newer commercial devices, especially smartphones released after 2017, to an external power monitor requires significant effort due to the increasing level of integration of their electronic circuits. Fig. \ref{fig:batteryconnector} compares the battery connector in an older Samsung smartphone, the Galaxy S5, released in 2014, with that of a newer Samsung model, the Galaxy S20, released in 2020. The battery connector in a smartphone is used to connect a battery to its integrated circuit board. Older smartphones, including the Galaxy S5, use a specific type of battery connector known as a "snap-type connector". Featuring four metal prongs, as shown in Fig. \ref{fig:S5}, the snap-type connector allows for easy identification of the positive and negative terminals and enables connection to an external power monitor. However, advanced smartphones such as the Galaxy S20 use a proprietary, tiny, and delicate Flexible Printed Circuit (FPC) battery connector, as shown in Fig. \ref{fig:S20}. The FPC connector's small size and delicate construction make it challenging to work with, requiring specialized tools and expertise to connect it to an external power monitor that offers higher accuracy. 
This might be one of the main reasons that recent research papers typically rely on the built-in current sensor for measuring coarse-grained power consumption on mobile and edge devices \cite{apicharttrisorn2019frugal, ran2018deepdecision, ignatov2018ai}.

\textit{Consequently, although external power monitors with high sampling rates show promising accuracy in measurement, the challenges associated with connecting newer commercial devices to such external monitors can be a significant barrier.}

\textbf{C2: Extensibility.}
In recent years, we have witnessed a significant surge in the development of DNNs, particularly those specifically designed to address the increasing demand for mobile and edge devices. This has led to the invention of several milestone Convolutional Neural Network (CNN) models, including, but not limited to, AlexNet, DenseNet, GoogleNet, and MobileNet.
Moreover, the advent of Neural Architecture Search (NAS) has accelerated advancements in the design and optimization of novel CNN models by automating the design process and facilitating customization.
While measuring the energy consumption of DNN inferences on real devices is highly desirable for various tasks, such as serving as a ground-truth dataset for training energy predictors for on-device deep learning, it is practically infeasible and excessively time-consuming to measure all DNN models individually. For example, we spend approximately $2.1$ days to measure 200 models on a single device, while ProxylessNAS \cite{cai2018proxylessnas} explores nearly 0.3 million models in a single round of search. 
This predicament leads to a critical challenge: how can we ensure the observations and measurements derived from a limited DNN model space can be extensible to new (unseen) DNN models?

\textit{Consequently, the huge and expansive model-design space significantly challenges the extensibility of energy measurements on real mobile and edge devices.}

\textbf{C3: Understandability.}
In addition to valuing research aimed at reducing the energy consumption of DNN executions, it is essential that our measurement study is accessible to a wide audience, such as end-users with non-technical backgrounds. 
As we presented in Section \ref{sc:Introduction}, end-users consider the energy efficiency of their devices as one of the most critical factors. Results that are easy to understand can help end-users make informed purchasing decisions. For instance, presenting an energy efficiency score, ranging from $0$ to $100$, could be more straightforward and easier to understand than simply telling the end-user that the device will consume $120.090$ mJ per inference to run MobileNetv1 with CPUs. Consequently, end-users can compare different devices and choose the one that best suits their needs. On the other hand, for the research community, an easily adoptable measurement method or energy dataset can accelerate progress in developing energy-efficient DNN models, designing energy predictors, or searching for DNN models with energy/power constraints within a vast model-design space. Currently, due to a lack of sustainability mindset, the optimization of DNNs is primarily driven by research focused on achieving high inference accuracy and minimizing latency. 

\textit{We hope our work can help shift the mindset of both end-users and the research community towards sustainability, a principle that drives our research.}

\begin{table*}
\begin{threeparttable}
\small
  \caption{Specifications of Measured Edge Devices and Chipsets}
  \vspace{-0.1in}
  \label{tab:phonespecs}
  \begin{tabular}{llcccccccc}
    \toprule
\multicolumn{2}{l}{\multirow{2}{*}{Model}} & OnePlus & Xiaomi & Huawei & Huawei & Huawei & Huawei & Xiaomi & Motorola\\
\multicolumn{2}{l}{}                       & 8 Pro & Redmi Note8 & Mate40 Pro & P40 Pro & P40 Lite & P40 Lite E & Redmi K30 Ultra & One Macro\\\midrule  
\multicolumn{2}{l}{SoC}                    & SD 865 & SD 665 & Kirin 9000 & Kirin 990 5G & Kirin 810 & Kirin 710F &  Dimensity1000+ & Helio P70 \\\midrule
\multicolumn{2}{l}{Vendor}                 & Qualcomm & Qualcomm & HiSilicon & HiSilicon & HiSilicon & HiSilicon & MediaTek & MediaTek\\\midrule

\multirow{3}{*}{CPU}                & M  & A77+A55 & A73+A53 & A77+A55 & A76+A55 & A76+A55 & A73+A53 & A77+A55 & A73+A53 \\
                                    & C1  & 4+4 & 4+4 & 4+4 & 4+4 & 2+6 & 4+4 & 4+4 & 4+4  \\
                                    & F1 & 2.84 GHz & 2.0 GHz & 3.13 GHz & 2.86 GHz & 2.27 GHz & 2.2 GHz & 2.6 GHz & 2.1 GHz\\\midrule
                                    
\multicolumn{2}{l}{GPU}            & Adreno 650 & Adreno 610 & Mali G78 & Mali G76 & Mali G52 & Mali G51 & Mali G77 & Mali G72\\\midrule

\multicolumn{2}{l}{Dedicated}      & Hexagon698 & Hexagon686 & Ascend Lite+Tiny & Lite+Tiny & D100Lite &  & MediaTek3.0 & MediaTek \\
\multicolumn{2}{l}{AI}             & DSP  & DSP & NPU & NPU & NPU & None & APU & APU \\
\multicolumn{2}{l}{accelerator}    &   &   & Da Vinci 2.0 & Da Vinci & Da Vinci & &\\\midrule

\multicolumn{2}{l}{OS (Android)}             & 10 & 10  & 10 & 10 & 10 & 10 & 10 &  9 \\\midrule
\multicolumn{2}{l}{NNAPI}          & \multirow{2}{*}{Yes} & \multirow{2}{*}{Yes} & \multirow{2}{*}{Yes}  & \multirow{2}{*}{Yes} & \multirow{2}{*}{Yes} & \multirow{2}{*}{Yes}  & \multirow{2}{*}{Yes} & \multirow{2}{*}{Yes} \\
\multicolumn{2}{l}{support}        &  &  &&&&&\\\midrule


\multirow{2}{*}{Battery}        & C2  & 4510 mAh & 4500 mAh  & 4400 mAh & 4200 mAh & 4200 mAh & 4000 mAh & 4500 mAh & 4000 mAh\\
                                & R & No & No & No & No & No & No & No & No \\\midrule
\multicolumn{2}{l}{Class}          & Flag  & Mid-range & Flag & Flag & Mid-range & Mid-range & Flag & Mid-range \\
\bottomrule
\end{tabular}
     \begin{tablenotes}
        \item[*] SD: Snapdragon, M: Microarchitecture, C1: CPU Cores, F1: Maximum Frequency, S: Display Size, C2: Battery Capacity, and R: If battery is removable.
    \end{tablenotes}
\end{threeparttable}
\end{table*}
\begin{figure*}[t]
\centering
\subfigure[OnePlus 8 Pro]
{\includegraphics[width=0.115\textwidth]{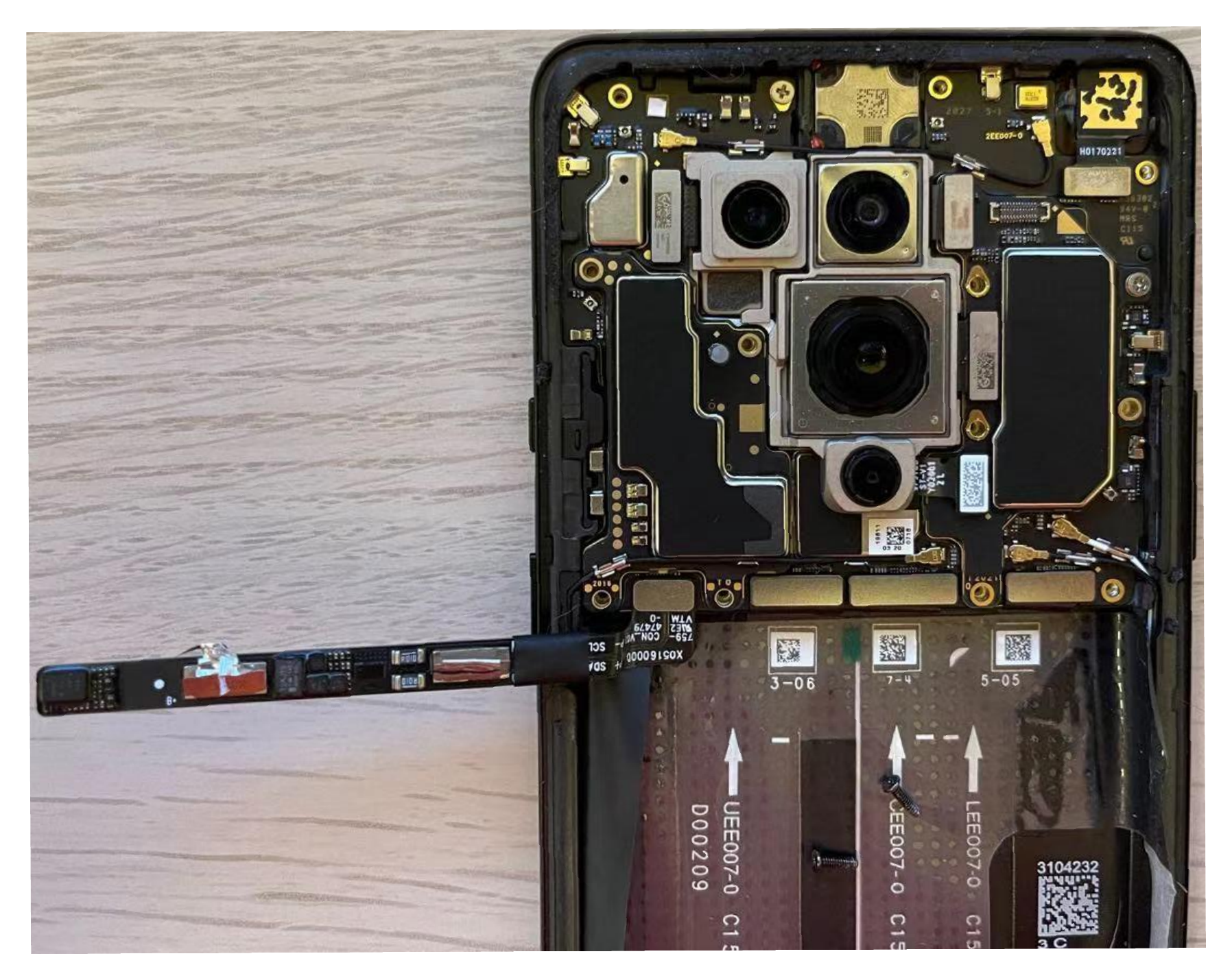}\label{fig:8Pro}}
\subfigure[Redmi Note8]
{\includegraphics[width=0.115\textwidth]{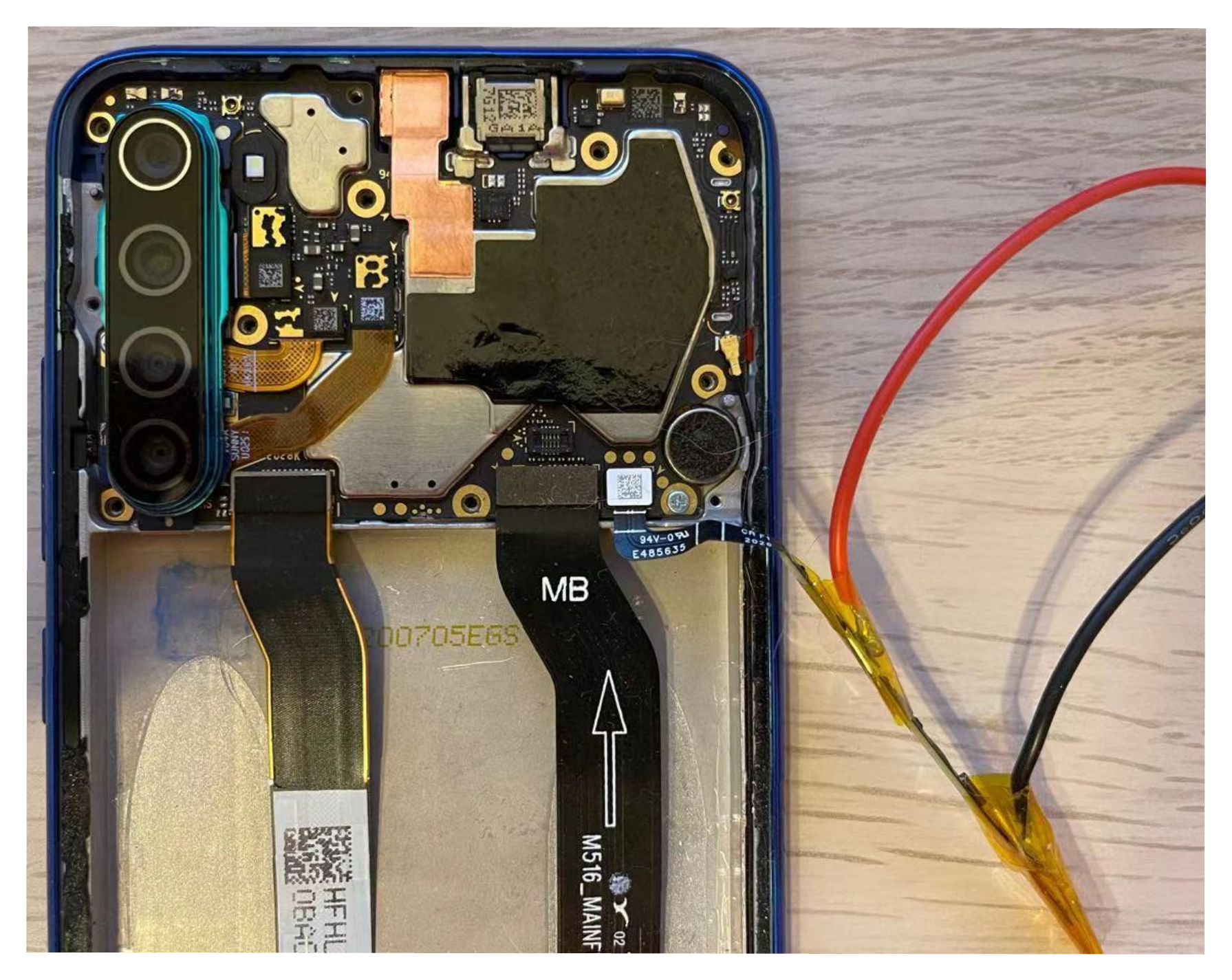}\label{fig:redminote8}}
\subfigure[Mate40 Pro]
{\includegraphics[width=0.115\textwidth]{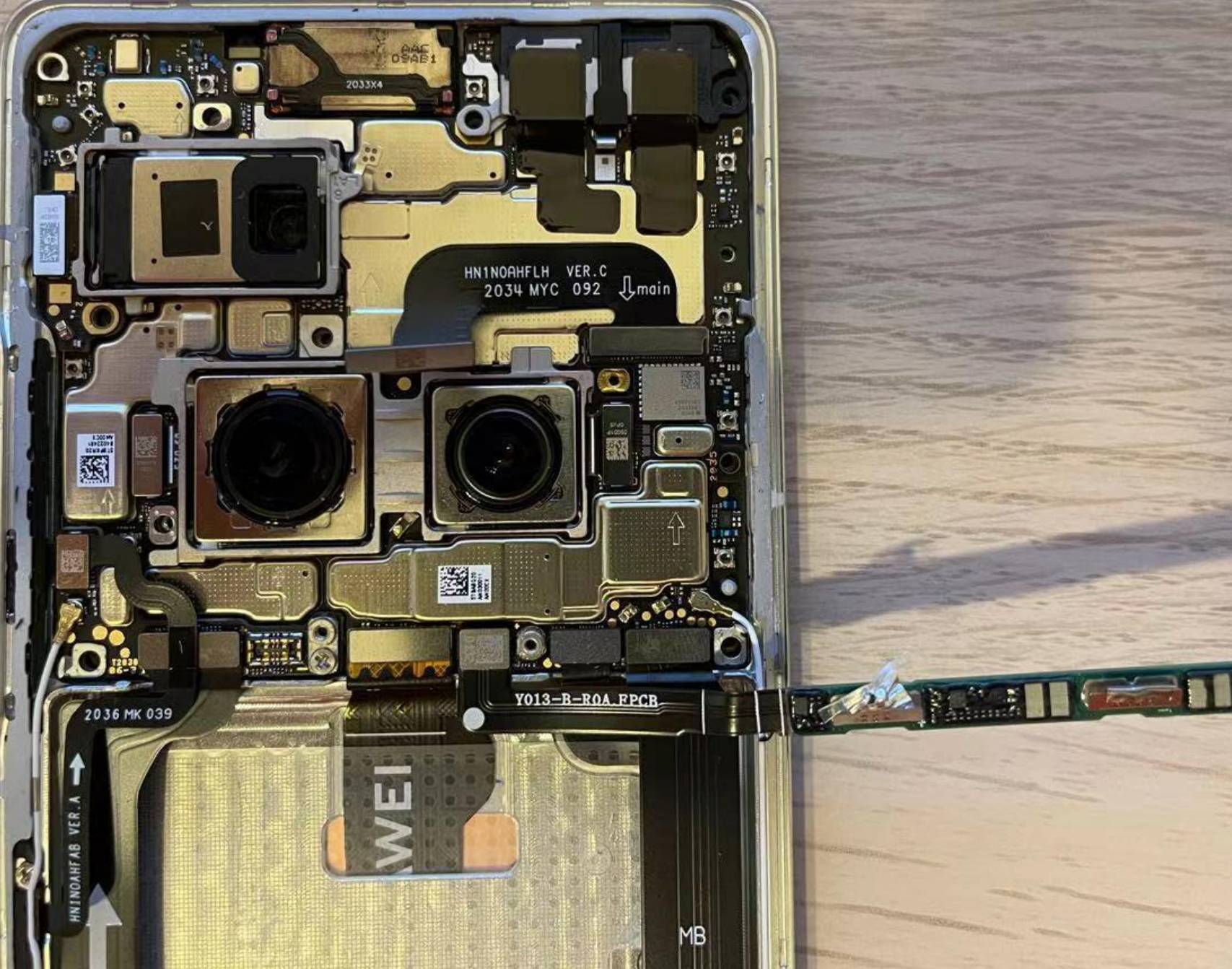}\label{fig:mate40}}
\subfigure[P40 Pro]
{\includegraphics[width=0.115\textwidth]{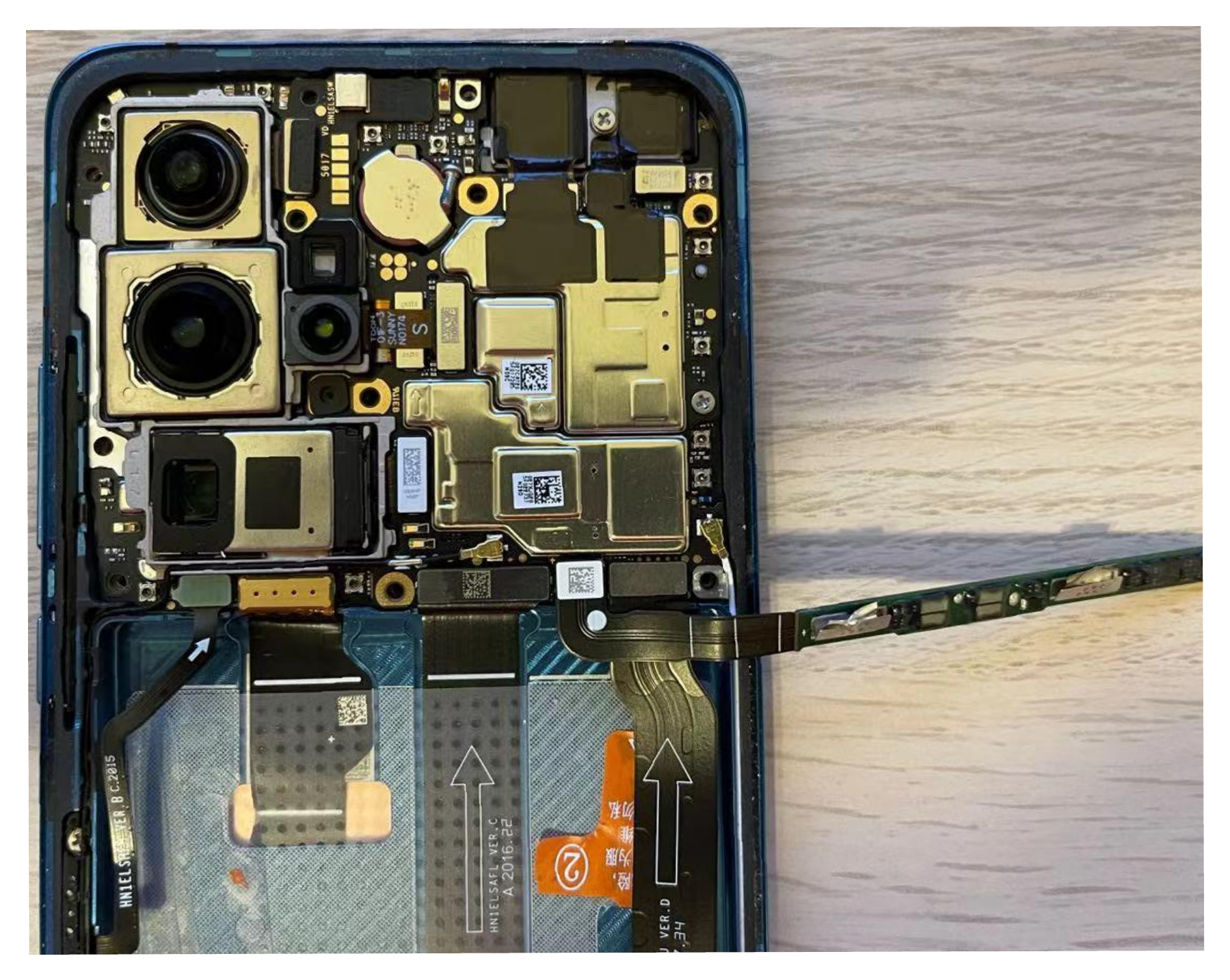}\label{fig:P40}}
\subfigure[P40 Lite]
{\includegraphics[width=0.115\textwidth]{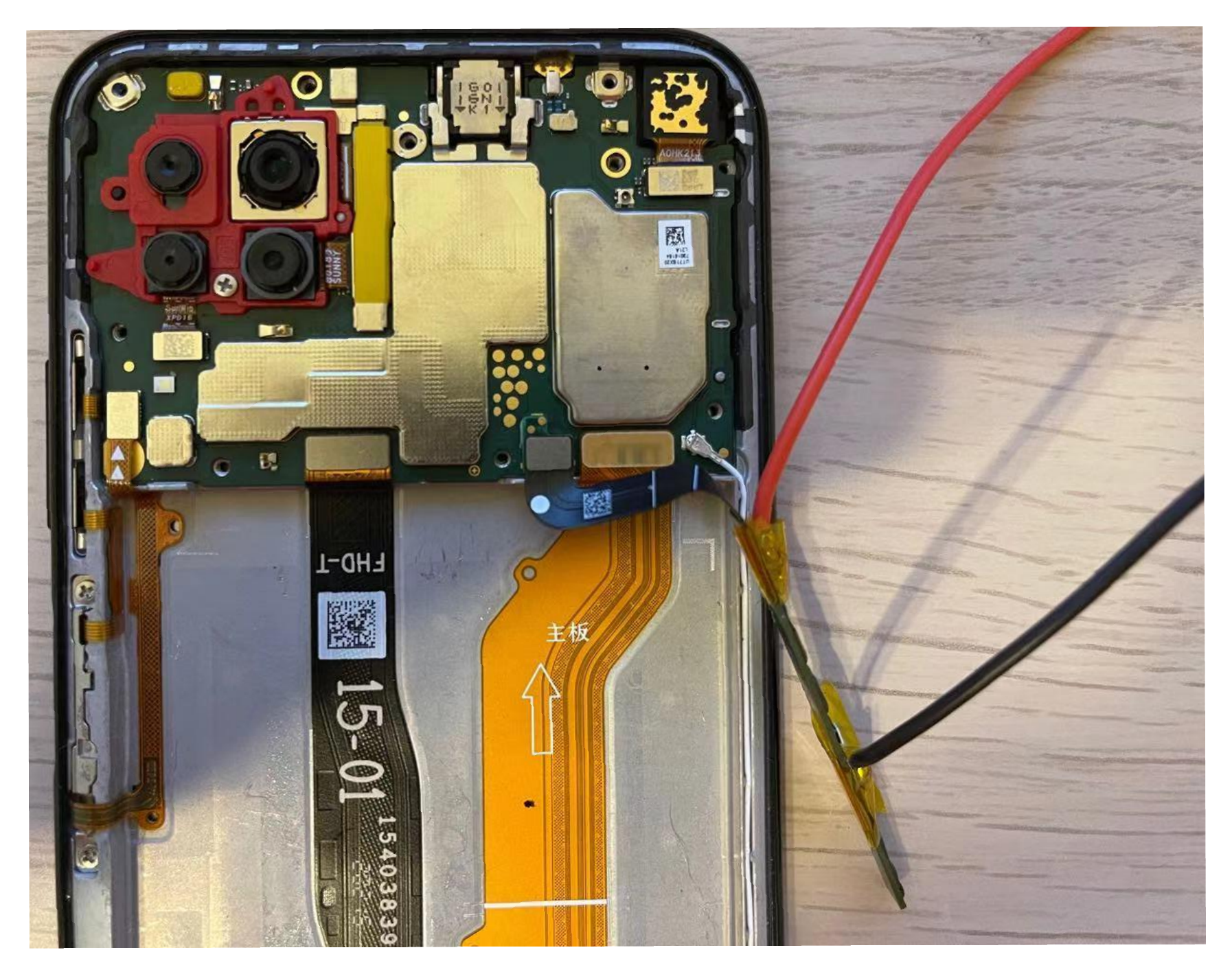}\label{fig:P40Lite}}
\subfigure[P40 Lite E]
{\includegraphics[width=0.115\textwidth]{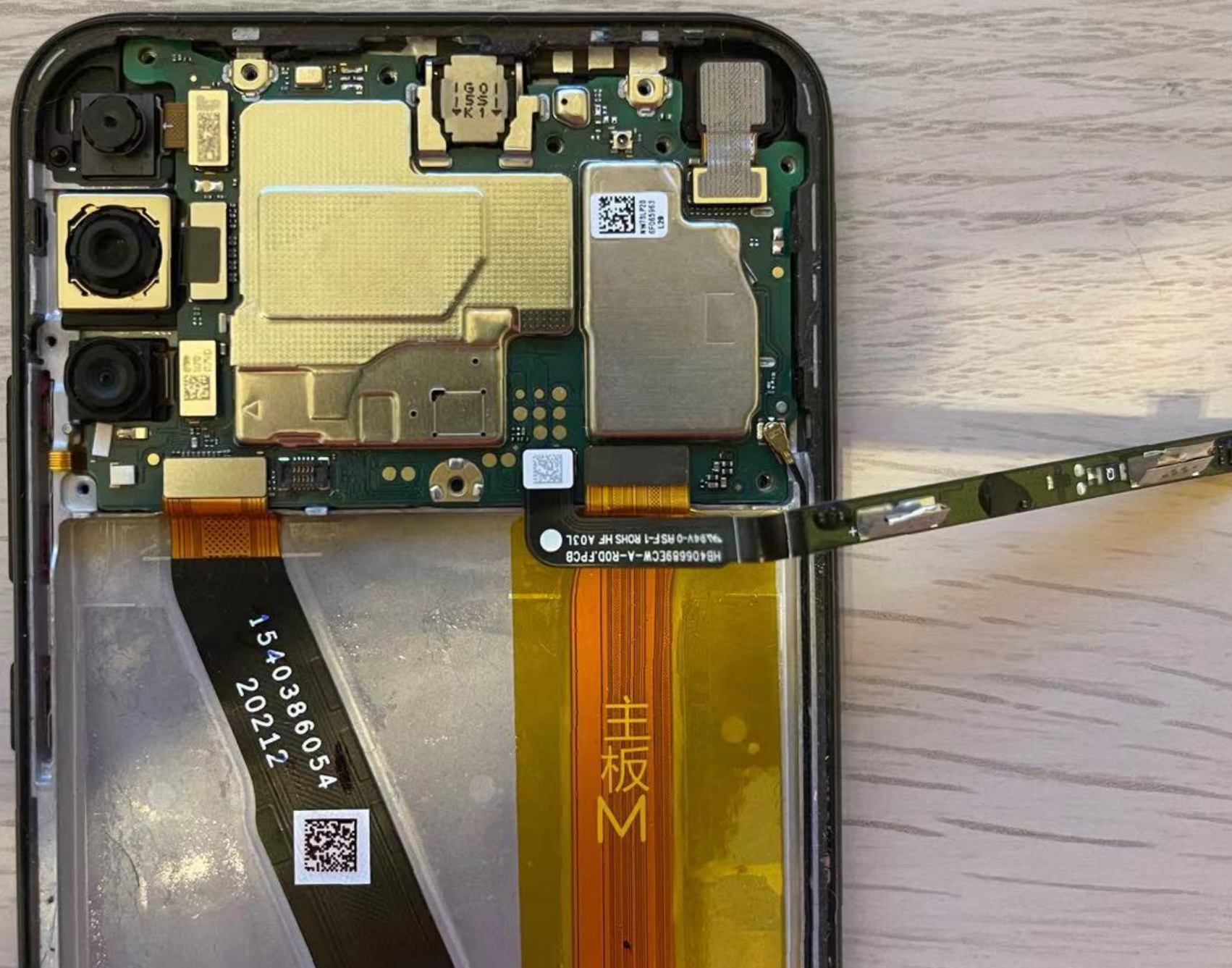}\label{fig:P40LiteE}}
\subfigure[K30 Ultra]
{\includegraphics[width=0.115\textwidth]{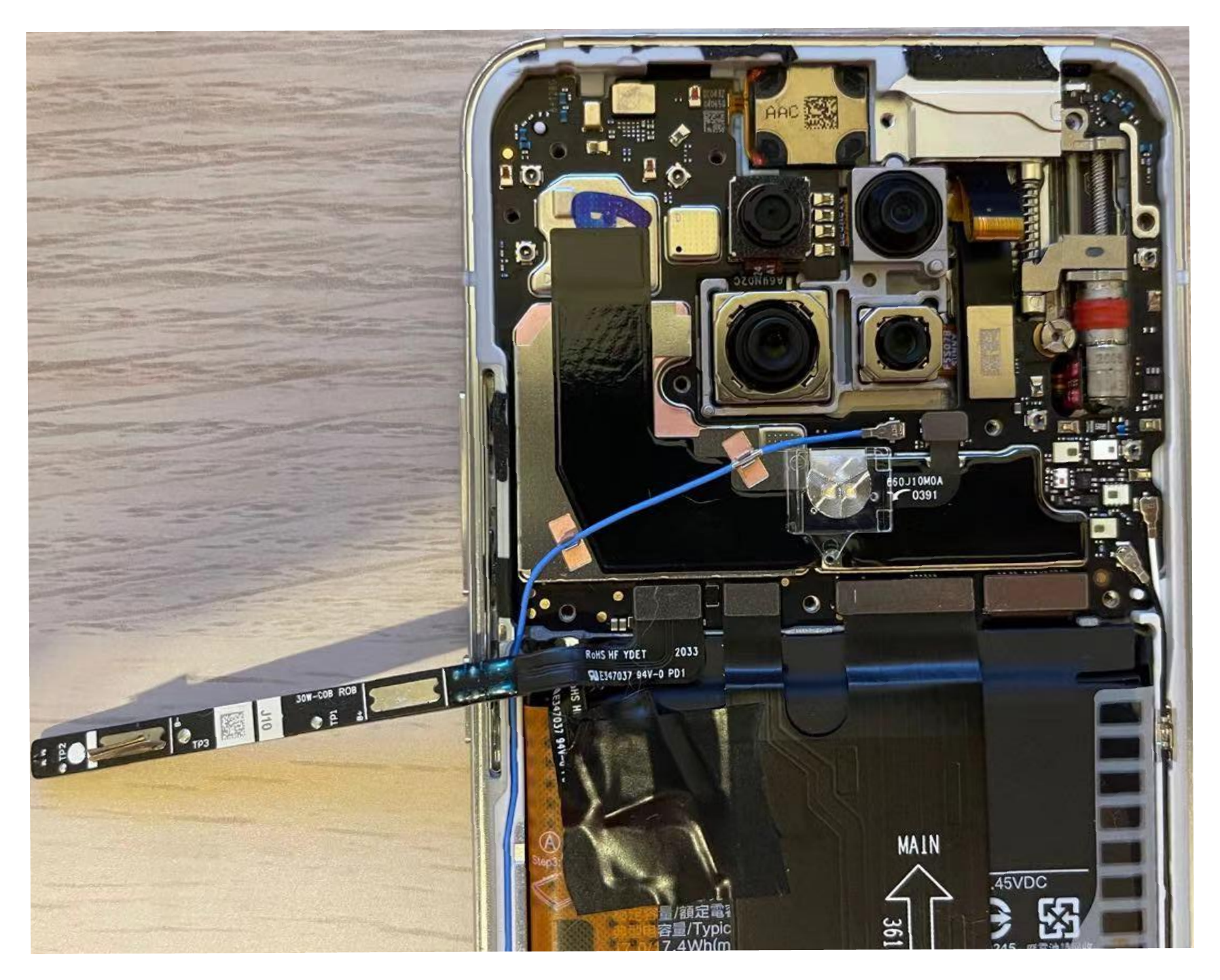}\label{fig:redmik30}}
\subfigure[One Macro]
{\includegraphics[width=0.115\textwidth]{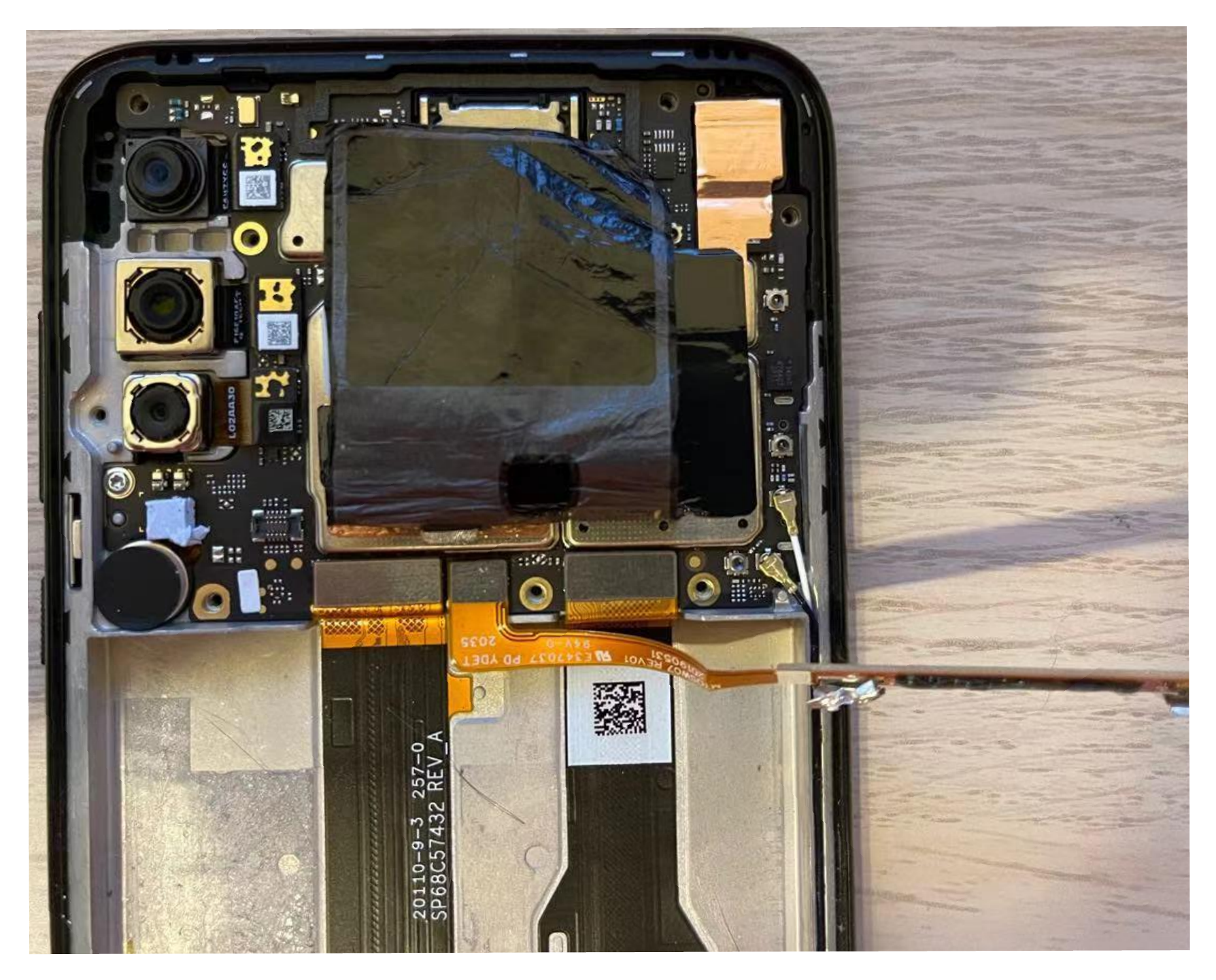}\label{fig:onemacio}}
\vspace{-0.2in}
\caption{Measured devices with segregated BMS chips.}
\label{fig:phoneteardown}   
\end{figure*}
\vspace{-0.05in}

\section{Energy Measurement and Dataset}
\label{sc:dataset}
We conduct a measurement study and create three energy datasets: kernel-, model-, and application-level datasets. Overall, we collect fine-grained power traces and accurate energy consumption data for (1) $16$ types of kernels with $1,847$ unique configurations, (2) nine SOTA DNN models with $50$ variants each, and (3) six widely-used edge AI applications on eight commercial edge devices.

\subsection{Energy Measurement}
We develop a reproducible energy measurement methodology, which facilitates the collection of accurate and fine-grained power consumption of kernels, DNN models, and end-to-end edge AI applications on modern edge devices.

\textbf{Proposed solution for C1: accuracy.} 
As discussed in Section \ref{sc:challenges}, although external power monitors demonstrate promising accuracy and time granularity for tracing power variations within a DNN execution, establishing a physical connection between a modern edge device with an FPC battery connector and such a monitor is nontrivial.
To address this challenge, we first use a mechanic mobile device DC power cable \cite{DCcable} that is designed to fit multiple device models, including those with FPC connectors, to connect the tested devices to an external power monitor. This method requires little effort on the part of the benchmarking researchers. However, we find that the tested devices cannot boot due to the lack of proprietary battery management system (BMS) chips. BMS is an electronic system that manages and monitors the performance and safety of a device battery, and is typically attached to the battery in modern edge devices. The device OS must communicate with the proprietary BMS to check the status and safety of the battery before allowing the phone to power on. Hence, the device cannot boot if its battery is disconnected or an unauthorized battery is connected. 
We have studied multiple alternatives to address this issue, and we find that the most effective method is to segregate the BMS chip from the device battery without tearing it down, and use it as a bridge to connect the device to the external power monitor. This method strikes a good balance between the effort required and reproducibility. We have validated this method on eight different modern smartphones, as illustrated in Fig. \ref{fig:phoneteardown}. All of the tested devices are able to power on with full functionality using this method.

\textit{We develope a detailed documentation to provide step-by-step instructions on how to implement this method on other modern edge devices, which will help the community to reproduce measurements and apply this technique to their own research.}

\textbf{Rules for measurement.}
Since the power consumption of mobile and edge devices can be easily influenced by the environment, such as heat dissipation and background activities, it is crucial to create specific rules for measurement. 
These rules can bolster the consistency and reliability of power measurements across diverse devices and testing conditions.
By controlling and accounting for environmental factors, we can mitigate their influence on our power data collection, and thus gain a more accurate understanding of the inherent power and energy consumption characteristics of DNN executions. To this end, we establish a set of rules for power measurements. Through our observation, these rules effectively ensure consistency and reproducibility\footnote{Although understanding how the power consumption of DNN executions may vary with noisy background activities is important (since it is close to practical use cases), it is equally crucial to isolate and understand the intrinsic power characteristics of the DNNs, independent of these variations. This is one of the primary goals of our measurement study in this paper.}.
\begin{itemize}
    \item Disable adaptive brightness and set the display to the lowest brightness level.
    \item Turn off WiFi, Bluetooth, cellular network, and Near-Field Communication (NFC) interfaces to minimize the interference on the accuracy of power measurements.
    \item Shut down and disable any background applications and services to minimize the interference on the accuracy of measurements.
    \item Conduct measurements with a room-temperature between $20$ and $25^\circ$C.
    \item Maintain an air gap with proper ventilation to regulate the temperature of the smartphone and prevent run-time thermal throttling.
    \item Configure the screen refresh rate to $60$ Hz.
    \item Configure the camera sample rate to 15 frames per second, if the executed edge AI applications require the use of the device camera.
    \item Set up a 2-minute cooldown interval between individual tests to allow the device to cooldown.
\end{itemize}

\textbf{Devices and tools.}
We select eight modern edge devices with distinct mobile SoCs that include at least one high-end and one mid-range SoC from leading chipset vendors, such as Qualcomm, HiSilicon, and MediaTek. Their specifications are presented in Table \ref{tab:phonespecs}. The selected mobile SoCs can serve as representative examples of advanced and widely used mobile AI silicons in the past two years. Unless stated, all power consumption data are measured by the Monsoon power monitor with a $5000$ Hz sampling rate. 
The latency of DNN inferences, including both model-level and kernel-level latencies, is measured by the TFLite \texttt{benchmark tool} \cite{TFbenchmark}.

\begin{table*}[t]
\centering
\caption{Measured kernels per device in our kernel-level dataset.}
\vspace{-0.1in}
\resizebox{\textwidth}{!}{
\footnotesize
\begin{tabular}{ccccccc}
\toprule
\multicolumn{1}{c}{\multirow{3}{*}{Kernels}} & \multicolumn{2}{l}{Energy Consumption (mJ)} & \multicolumn{2}{c}{\# Measured kernels} & \multicolumn{1}{c}{\multirow{2}{*}{Avg. FLOPs}} & \multicolumn{1}{c}{\multirow{3}{*}{Configurations}} \\ 
\cline{2-5}
    & CPU & GPU        & \multirow{2}{*}{CPU} & \multirow{2}{*}{GPU} &   \multirow{2}{*}{(M)}    \\
    & min - max & min - max &  
    &       &      \\                              
\hline
\texttt{conv$\doubleplus$bn$\doubleplus$relu}    & 0.002 - 1200.083  & 0.002 - 120.152  & 1032 & 1032 & 250.137 &   ($HW, C_{in}, C_{out}, KS, S$)    \\
\texttt{dwconv$\doubleplus$bn$\doubleplus$relu}  & 0.022 - 222.609   & 0.016 - 0.658    & 349 & 349 &  28.364 &      ($HW, C_{in}, KS, S$)    \\ 
\texttt{bn$\doubleplus$relu}                     & 0.002 - 161.334   & 0.001 - 14.594   & 100 & 100 & 4.710 &    ($HW, C_{in}$)    \\ 
\texttt{relu}                                    & 0.001 - 141.029   & 0.003 - 6.86     & 46 & 46 & 7.983 &  ($HW, C_{in}$)    \\ 
\texttt{avgpool}                                 & 0.066 - 7.711     & 0.034 - 1.142    & 28 & 28 & 0.670 &  ($HW, C_{in}, KS, S$)    \\ 
\texttt{maxpool}                                 & 0.054 - 7.779     & 0.032 - 1.214    & 28 & 28 & 0.521 &  ($HW, C_{in}, KS, S$)    \\ 
\texttt{fc}                                      & 0.038 - 94.639    & -                & 24 & - & 14.744 &  ($C_{in}, C_{out}$)    \\
\texttt{concat}                                  & 0.001 - 42.826    & 0.066 - 3.428    & 142 & 142 &  0 &    ($HW, C_{in1}, C_{in2}, C_{in3}, C_{in4}$)    \\ 
\texttt{others}                                  & 0.001 - 132.861   & 0.003 - 10.163   & 98 & 72 & - &    ($HW, C_{in}$)    \\ 
\bottomrule

\end{tabular}}
\label{tb:ker_data}
\end{table*}

\begin{table}[t]
\centering
\caption{Energy consumption of \texttt{conv$\doubleplus$bn$\doubleplus$relu} kernels with different configurations on mobile CPU.}
\vspace{-0.1in}
\small
\begin{tabular}{ccc}
\toprule
\multirow{2}{*}{} & \multicolumn{2}{c}{($HW, C_{in}, C_{out}, KS, S$)} \\ 
\cline{2-3}
                  & (112, 64, 128, 3, 1)   &    (28, 22, 22, 1, 1)       \\ 
\hline 
 Energy (mJ)           & 125.232              &    0.064    \\  
\bottomrule
\end{tabular}
\label{tb:ker_cp}
\end{table}

\subsection{Energy Dataset}
\textbf{Kernel-level.} 
As we introduced in Section \ref{sc:challenges}, kernels constitute the fundamental units of execution in deep learning frameworks, with their types and configuration parameters significantly influencing the energy consumption during DNN executions.
Table \ref{tb:ker_data} illustrates that \texttt{conv$\doubleplus$bn$\doubleplus$relu} kernels typically consume more energy than other kernel types.
Furthermore, the configuration for each kernel type varies. For \texttt{conv$\doubleplus$bn$\doubleplus$relu} and \texttt{dwconv$\doubleplus$bn$\doubleplus$relu} kernels, the primary configurations includes input height and width ($HW$)\footnote{In CNN models, input height usually is equal to input width.}, input channel number ($C_{in}$), output channel number ($C_{out}$), kernel size ($KS$), and stride ($S$). 
Table \ref{tb:ker_cp} presents a comparison of the energy consumption between two \texttt{conv$\doubleplus$bn$\doubleplus$relu} kernels with different configurations, both run on a mobile CPU. One kernel configuration consumes a considerable $125.232$mJ of energy, whereas the other expends a mere $0.064$mJ. As a result, examining the impact of kernel configurations on energy consumption lays the foundation for a comprehensive understanding of energy consumption during DNN executions on edge devices.

To this end, we present our kernel-level energy dataset collected from real edge devices. To build the dataset, as presented in Table \ref{tb:ker_data}, we initially generate a large number of kernels with a variety of types (16 types for CPU and 10 types for GPU) featuring a range of configurations in the \texttt{tflite} format (e.g., $1032$ \texttt{conv$\doubleplus$bn$\doubleplus$relu} and $349$ \texttt{dwconv$\doubleplus$bn$\doubleplus$relu} kernels). The number of sampled configurations for each kernel type hinges on two main factors: its configuration dimension and its impact on the overall energy consumption during DNN executions (e.g., we observe that the \texttt{conv$\doubleplus$bn$\doubleplus$relu} kernel accounts for more than $70\%$ of the total energy consumption in most SOTA CNN models on edge devices). These kernel configurations are randomly sampled in accordance with the sampling strategy proposed in \cite{zhang2021nn}.
Then, we measure the average power consumption and inference latency for each generated kernel running on individual edge devices. Each power and latency value is the average of at least $100$ inference runs. We conduct these measurements independently on both CPUs and GPUs.
As shown in Table \ref{tb:ker_data}, our kernel-level energy dataset spans a broad spectrum with different levels of energy consumption. 

\begin{figure*}[htb]
\centering
\subfigure[$C_{in}$/$C_{out}$/$S$=$20$/$120$/$1$]
{\includegraphics[width=0.2\textwidth]{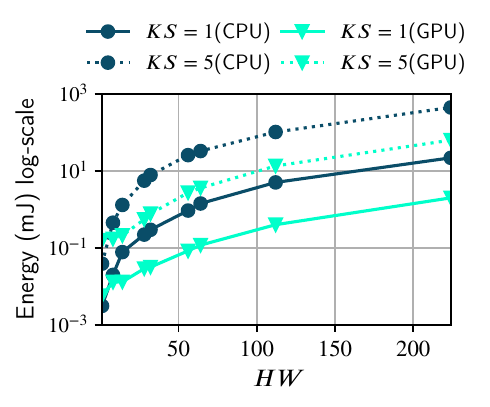}\label{fig:cpuhw}}
\subfigure[$C_{in}$/$C_{out}$/$S$=$20$/$120$/$1$]
{\includegraphics[width=0.2\textwidth]{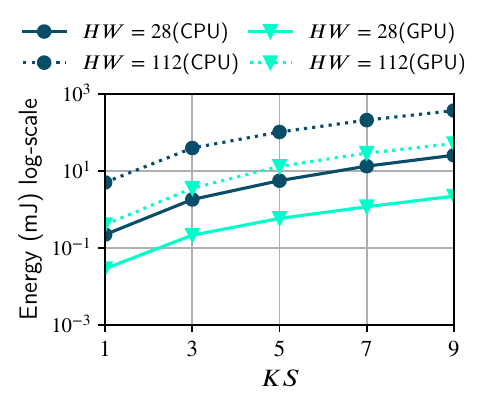}\label{fig:cpuks}}
\subfigure[$HW$/$C_{out}$/$S$=$28$/$120$/$1$]
{\includegraphics[width=0.19\textwidth]{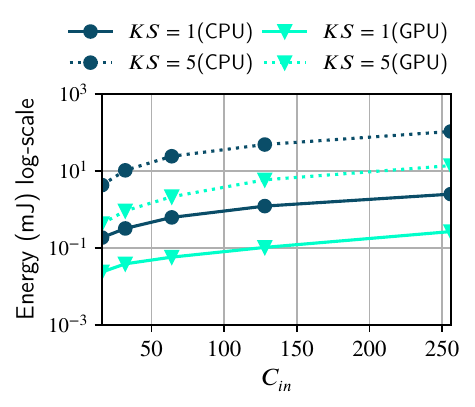}\label{fig:cpucin}}
\subfigure[$HW$/$C_{in}$/$S$=$28$/$20$/$1$]
{\includegraphics[width=0.19\textwidth]{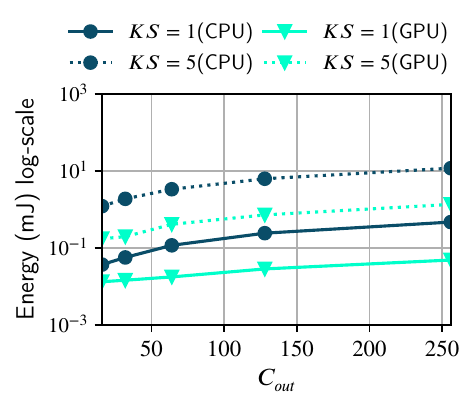}\label{fig:cpucout}}
\subfigure[$HW$/$C_{in}$/$C_{out}$=$28$/$20$/$120$]
{\includegraphics[width=0.19\textwidth]{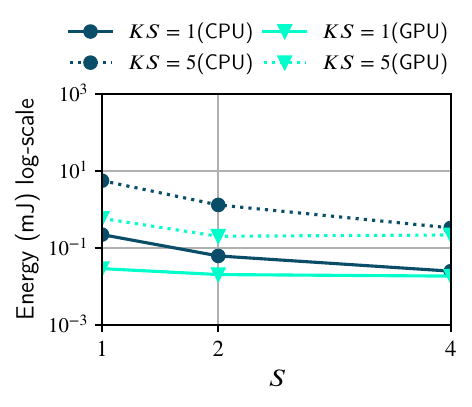}\label{fig:cpus}}
\vspace{-0.15in}
\caption{Energy consumption of \texttt{conv$\doubleplus$bn$\doubleplus$relu} vs. kernel configurations.}
\label{fig:conv_config}   
\end{figure*}

In Fig. \ref{fig:conv_config}, we seek to investigate how the five configurations (i.e., $HW$, $C_{in}$, $C_{out}$, $KS$, and $S$) impact the energy consumption of \texttt{conv$\doubleplus$bn$\doubleplus$relu}. In each evaluation, we vary a single configuration (e.g., $HW$) while maintaining the other four constants. The results reveal that the relationship between the energy consumption and the configurations is non-linear. As illustrated in Fig. \ref{fig:cpuhw}, the energy consumption demonstrates a progressive increase with the growth of $HW$. For instance, when running on the mobile CPU, the energy consumption of \texttt{conv$\doubleplus$bn$\doubleplus$relu} increases by approximately $1.85\times$ (0.077mJ to 0.22mJ), $3.2\times$ (0.22mJ to 0.93mJ), $4.37\times$ (from 0.93mJ to 5.0mJ), $3.36\times$ (5.0mJ to 21.81mJ), as $HW$ doubles from 14 to 28, 28 to 56, 56 to 112, and 112 to 224, respectively.
While operating on the mobile GPU, the energy consumption of the \texttt{conv$\doubleplus$bn$\doubleplus$relu} exhibits a similar trend but at a different rate. In this case, its energy consumption increases by roughly $1.21\times$ (0.013mJ to 0.029mJ), $1.89\times$ (0.029mJ to 0.083mJ), $3.79\times$ (0.083mJ to 0.399mJ), $3.98\times$ (0.399mJ to 1.988mJ) when $HW$ doubles from 14 to 28, 28 to 56, 56 to 112, and 112 to 224, respectively.
Moreover, we find that $KS$ has the most significant impact on the energy consumption of \texttt{conv$\doubleplus$bn$\doubleplus$relu}. 
This is because the majority of energy consumption of kernel \texttt{conv$\doubleplus$bn$\doubleplus$relu} is attributed to convolutional layer. Within the convolutional layer, $KS$ has the most significant impact due to its quadratic relationship with computational cost, while other parameters have a linear relationship.
Specifically, when doubling each of the configuration, $KS$ (from 3 to 5), $HW$ (from 14 to 28), $C_{in}$ (from 128 to 256), and $C_{out}$ (from 128 to 256), the corresponding increases in energy consumption are approximately $2.08\times$, $1.85\times$, $1.05\times$, and $1.18\times$ respectively. This finding demonstrates the disproportionate influence of $KS$ on energy consumption relative to the other parameters.

\textit{Insights: the above observations underscore the importance of adaptive configuration selection in enhancing the energy efficiency of DNNs on edge devices.  Given that our kernel-level dataset covers a wide range of configurations, each associated with an energy consumption label, it can serve as a valuable resource for guiding the selection of optimal configurations, searching for energy-efficient kernel configurations that meet specific energy constraints, and training kernel-level energy predictors (present in Section \ref{sc:pred}).}

\begin{figure*}[t]
\centerline{\includegraphics[width=1.0\textwidth]{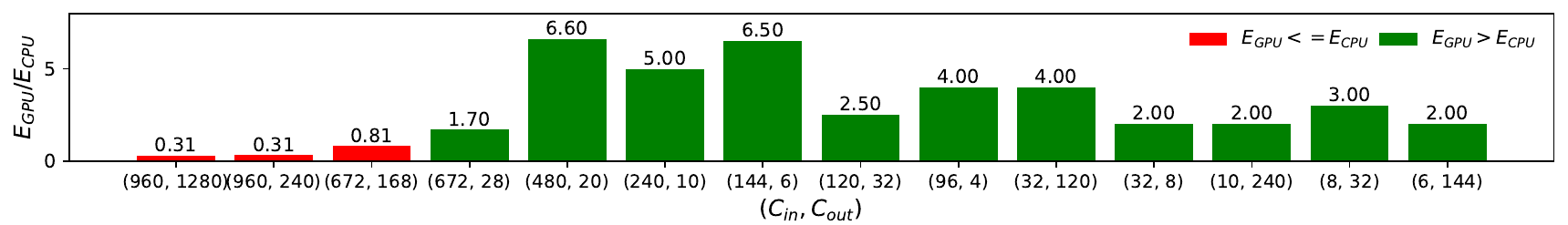}}
\vspace{-0.15in}
\caption{Comparison of energy consumption of \texttt{conv$\doubleplus$bn$\doubleplus$relu} with identical configurations on mobile CPU and GPU ($HW = 1, KS = 1, S = 1$, measured device: Huwei P40 Pro). Using the mobile GPU to execute the kernel does not always save the device's energy compared to using the mobile CPU.}
\label{fig:gpuvcpu}
\end{figure*}

\begin{figure}[t]
\centering
\subfigure[HiSilicon Kirin 990 5G (4 $\times$ CortexA76 $+$ 4 $\times$ CortexA55 CPUs)]
{\includegraphics[width=0.48\textwidth]{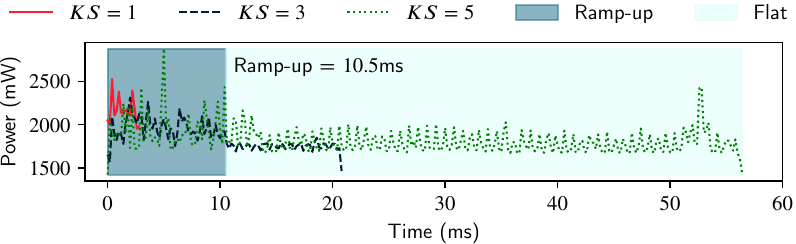}\label{fig:k990}}
\subfigure[HiSilicon Kirin 810 (2 $\times$ CortexA76 $+$ 6 $\times$ CortexA55 CPUs)]
{\includegraphics[width=0.48\textwidth]{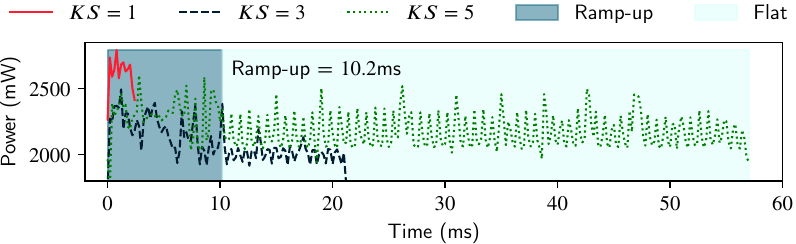}\label{fig:k810}}
\subfigure[Qualcomm Snapdragon 855 (4 $\times$ CortexA76 $+$ 4 $\times$ CortexA55 CPUs)]
{\includegraphics[width=0.48\textwidth]{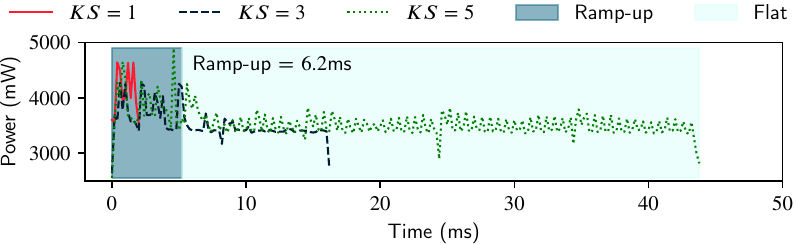}\label{fig:sd855}}
\vspace{-0.2in}
\caption{Measured fine-grained power slices for \texttt{conv$\doubleplus$bn$\doubleplus$relu} with $HW = 112, C_{in} = 20, C_{out} = 120, S = 1$. The intra-kernel power variation exhibits a ``high-initial, flat-later" pattern.}
\label{fig:comp_power}   
\end{figure}

Fig. \ref{fig:gpuvcpu} presents a comparison of the energy consumption between running \texttt{conv$\doubleplus$bn$\doubleplus$relu} with identical configurations on an edge device's CPU and GPU. Interestingly, we find that using the mobile GPU for executing the \texttt{conv$\doubleplus$bn$\doubleplus$relu} kernel does not always result in energy savings when compared to running the same kernel on a mobile CPU, especially when the $HW$ and $KS$ parameters are on the lower side. For instance, when testing on the Huawei P40 Pro with the kernel configurations of $HW = 1$, $KS = 1$, $C_{in} = 480$, and $C_{out} = 20$, we find that the energy consumed by the GPU exceeds that of the CPU by more than a factor of $6.6$. While the magnitude of this difference may vary across different edge devices, the overall pattern of increased energy consumption on the GPU under these conditions appears to be consistent.

\textit{Insights: This observation is crucial for designing effective kernel execution scheduling strategies on edge devices. Rather than only considering the type of kernel, the specific configuration of the kernel should also be taken into account when deciding where to execute it (e.g., on the mobile CPU or GPU).}

In addition, our kernel-level dataset includes fine-grained power traces for each individual kernel, referred to as \textit{power slices} in this paper. These collected power slices provide valuable insights for analyzing intra-kernel power variations.
One of the primary observations in power slices is that the intra-kernel power variation exhibits a ``high-initial, flat-later" pattern, illustrated in Fig. \ref{fig:comp_power}, when the kernel is executed on a mobile CPU and the execution time exceeds a certain threshold. 
Fig. \ref{fig:k990} reveals an initial power surge at the beginning of kernel execution on Huawei P40 Pro, equipped with a Kirin 990 5G chipset.
This ramp-up phase continues for approximately $10.5$ms. Following the initial ramp-up, the power consumption settles into a more consistent, flatter profile that persists until the end of the kernel's execution. 
We conduct validations across varying kernel configurations, with varying execution time, as well as on various edge devices to ascertain the consistency of this observation.
We find the same pattern on the measured devices, as demonstrated in Figs. \ref{fig:k810} and \ref{fig:sd855}. Interestingly, devices powered by chipsets from the same vendor (e.g., Kirin 990 5G and Kirin 810) exhibit a nearly identical ramp-up time (10.5ms and 10.2ms), while the Snapdragon 855's (this device is not listed in Table \ref{tab:phonespecs}) ramp-up time is around 6.2ms.
The ``high-initial, flat-later" pattern primarily arises due to power management techniques implemented in modern processors on edge devices.
For instance, the Dynamic Voltage and Frequency Scaling (DVFS) technique can dynamically adjust a processor's voltage and frequency during runtime, based on computational demands. At the beginning of a computationally intensive kernel execution, DVFS may increase the frequency to ensure the task's timely completion. It then lowers the frequency once the task becomes more manageable, resulting in a relatively flat power consumption profile.
The variation in ramp-up times among different chipsets and vendors can be attributed to the unique DVFS strategies they employ.

\textit{Insights: The ramp-up time can negatively impact power and energy efficiency on edge devices, particularly when executing kernels with relatively small configurations (where their execution time are less than the ramp-up time). 
As illustrated in Fig. \ref{fig:k810}, the ramp-up phase causes the \texttt{conv$\doubleplus$bn$\doubleplus$relu} kernel with $HW=112$, $C_{in}=20$, $C_{out}=120$, $S=1$ to consume $25.3\%$ and $19.6\%$ more power than kernels with larger $KS$s, specifically 3 and 5.
Nevertheless, kernels with smaller configurations are often preferred for implementation on edge devices to save computational resources. This highlights the critical importance of optimizing the ramp-up phase for edge devices. For instance, if the aforementioned kernel with $KS=1$ can be executed directly in the flat phase, it can result in a reduction of energy consumption by $23.1\%$.}


\begin{figure*}[t]
\centering
\subfigure[Mobile CPU]
{\includegraphics[width=0.49\textwidth]{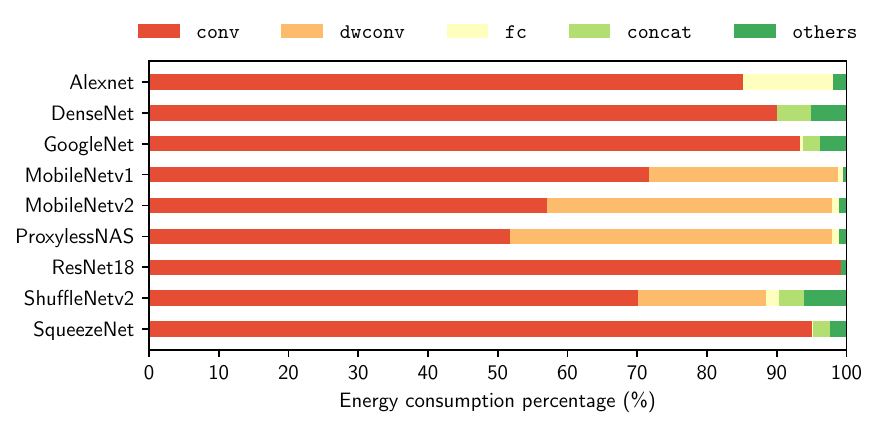}\label{fig:percpu}}
\subfigure[Mobile GPU]
{\includegraphics[width=0.49\textwidth]{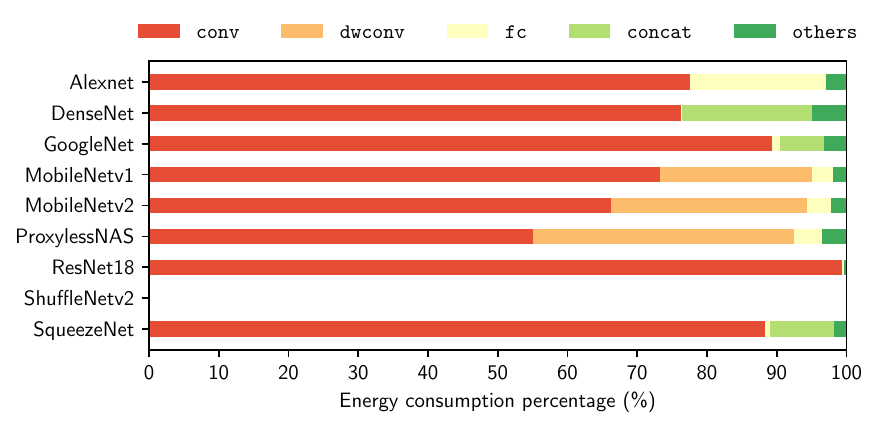}\label{fig:pergpu}}
\vspace{-0.15in}
\caption{DNN model energy consumption percentage breakdown. The top four most energy-consuming kernel types are \texttt{conv$\doubleplus$bn$\doubleplus$relu} (\texttt{conv}), \texttt{dwconv$\doubleplus$bn$\doubleplus$relu} (\texttt{dwconv}), \texttt{fc}, and \texttt{concat}.}
\label{fig:energyper}   
\end{figure*}

\textbf{Model-level.} We introduce our model-level energy dataset, which collects nine SOTA DNN models. These models represent a mix of both manually-designed and NAS-derived models, each with distinct kernel types and configurations. For each model, we generate $50$ variants for conducting power and energy measurements by re-sampling the $C_{out}$ and $KS$ for each layer. Specifically, we randomly sample the new output channel number from a range of $20\%$ to $180\%$ of the original $C_{out}$, while the $KS$ is sampled from the set of values: $\{1, 3, 5, 7, 9\}$. Table \ref{tb:mdl_data} presents the details of the measured DNN models. In general, running these models on mobile GPUs results in an energy consumption reduction of approximately $49\%$ to $79\%$, compared to the execution on mobile CPUs.

Fig. \ref{fig:energyper} presents the energy consumption breakdown of individual models by kernel types.  
The four kernel types that consume the most energy are \texttt{conv$\doubleplus$bn$\doubleplus$relu}, \texttt{dwconv$\doubleplus$bn$\doubleplus$relu}, \texttt{fc}, and \texttt{concat}. They account for $79.27\%$, $14.79\%$, $2.03\%$, and $1.5\%$ of the total model energy consumption on mobile CPUs, respectively. On mobile GPUs, these kernels represent $78.17\%$, $10.91\%$, $4.01\%$, and $4.28\%$ of the total model energy consumption.
Furthermore, in most models, \texttt{conv$\doubleplus$bn$\doubleplus$relu} and \texttt{dwconv$\doubleplus$bn$\doubleplus$relu} account for the main energy percentages. On average, \texttt{conv$\doubleplus$bn$\doubleplus$relu} and \texttt{dwconv$\doubleplus$bn$\doubleplus$relu} take $93.97\%$ and $87.74\%$ of the total model energy consumption on the mobile CPU and GPU, respectively. 

\begin{figure*}[t]
\centering
\subfigure[Power slice of AlexNet 1]
{\includegraphics[width=1.0\textwidth]{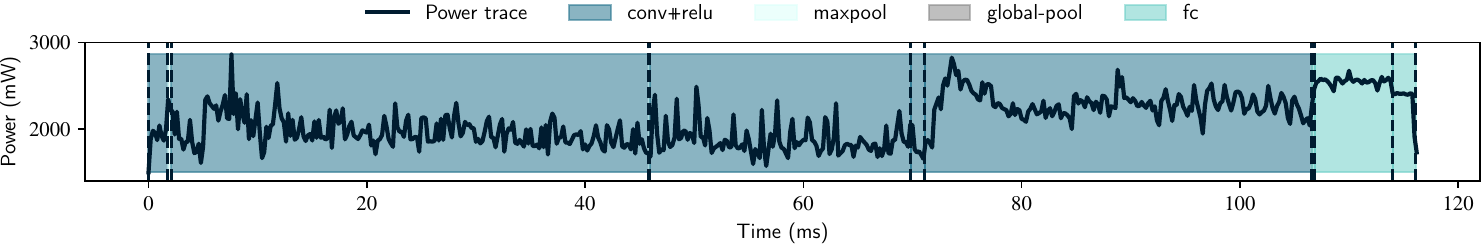}\label{fig:alex131}}
\subfigure[Power slice of AlexNet 2]
{\includegraphics[width=1.0\textwidth]{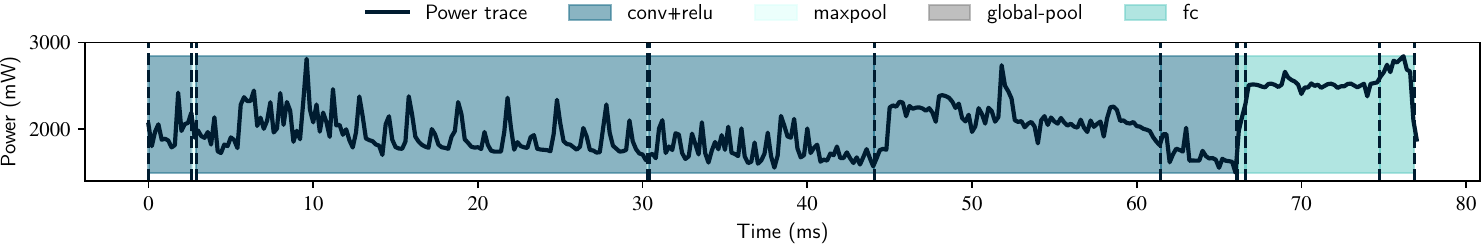}\label{fig:alex171}}
\vspace{-0.15in}
\caption{The model-level fine-grained power slices provided by our dataset can offer (1) a holistic view of the precise power variations associated with each kernel within the DNN model, (2) the temporal and sequential aspects of kernel executions, and (3) a visual approach to easily identify the power and energy bottlenecks within a specific DNN model.}
\label{fig:alex}   
\end{figure*}

In addition, our model-level dataset collects fine-grained power slices for all the measured DNN models. For instance, Fig. \ref{fig:alex} illustrates the measured power slices of two AlexNets with distinct kernel configurations, whose specific configurations are detailed in Table \ref{tb:alextable}. These model-level power slices offer (1) a holistic view of the precise power variations associated with each kernel within the DNN model, (2) the temporal and sequential aspects of kernel executions, and (3) a visual approach to easily identify the power and energy bottlenecks within a specific DNN model.

\begin{figure*}[t]
\centering
\subfigure[End-to-end processing pipeline for object detection and classification]
{\includegraphics[width=0.48\textwidth]{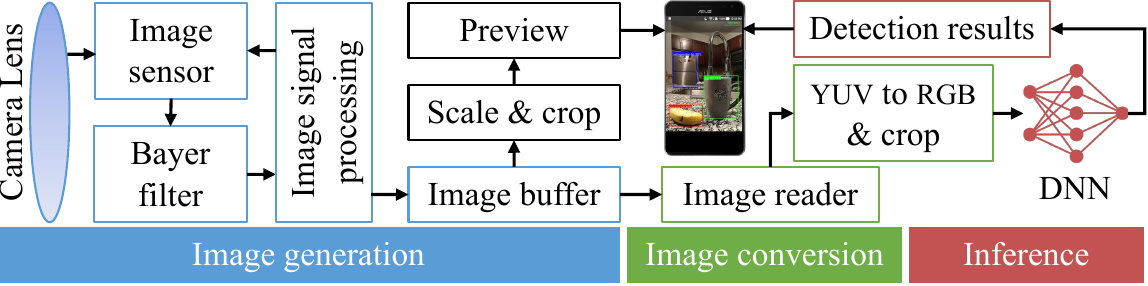}\label{fig:pipline}}
\subfigure[Energy consumption percentage breakdown]
{\includegraphics[width=0.46\textwidth]{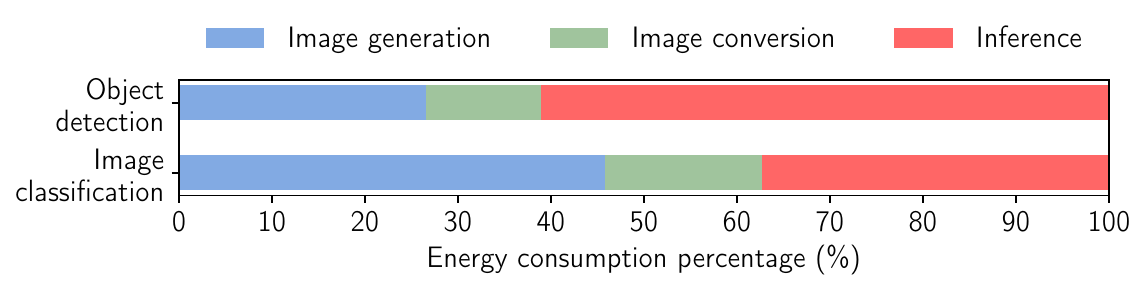}\label{fig:breakdown}}
\vspace{-0.15in}
\caption{End-to-end energy consumption breakdown for object detection and classification (application-level dataset).}
\label{fig:e2e}   
\end{figure*}

\begin{table}[t]
\centering
\caption{Measured DNN models in our model-level dataset.}
\vspace{-0.1in}
\small
\begin{tabular}{cccc}
\toprule
\multirow{3}{*}{Models} & \multicolumn{2}{c}{Energy consumption (mJ)} &  \multirow{2}{*}{Avg. FLOPs} \\
            & CPU       & GPU       &                  \\
            & min - max & min - max &            (M)   \\\hline
            
AlexNets    & 36.97 - 355.58 & 7.69 - 91.80 &  815 \\
DenseNets    & 231.93 - 488.87 & 66.21 - 133.58  & 1760  \\
GoogleNets    & 145.03 - 262.45 & 52.66 - 90.04  & 1535 \\
MobileNetv1s    & 53.59 - 136.79 & 17.36 - 42.44 & 519 \\
MobileNetv2s    & 30.85 - 175.07 & 8.81 - 48.35  &  419 \\
ProxylessNASs    & 58.34 - 162.11 & 17.70 - 49.29 & 526\\
ResNet18s    & 251.52 - 1432.67 & 64.19 - 391.97 &  3888 \\
ShuffleNetv2s    & 25.26 - 81.41 & - & 319 \\
SqueezeNets    & 92.55 - 388.16 & 34.55 - 134.65 &  1486 \\
\bottomrule
\end{tabular}
\label{tb:mdl_data}
\end{table}

\begin{table}[t]
\centering
\caption{Kernel configurations of two AlexNets.}
\vspace{-0.1in}
\small
\begin{tabular}{ccc}
\toprule
\multirow{3}{*}{Kernels} & \multicolumn{2}{c}{Configurations}\\
\cline{2-3}
                                    & AlexNet 1             & AlexNet 2 \\
                                    & (Fig. \ref{fig:alex131}) & (Fig \ref{fig:alex171}) \\
\hline    
\texttt{conv$\doubleplus$relu 1}    & (224, 3, 89, 5, 4)    & (224, 3, 70, 7, 4)    \\
\texttt{maxpool 1}                  & (224, 89, 3, 2)       & (55, 70, 3, 2) \\
\texttt{conv$\doubleplus$relu 2}    & (28, 89, 153, 7, 1)   & (28, 70, 115, 7, 1) \\
\texttt{maxpool 2}                  & (28, 153, 3, 2)       & (28, 115, 3, 2) \\
\texttt{conv$\doubleplus$relu 3}    & (13, 153, 460, 5, 1)  & (13, 115, 345, 5, 1) \\
\texttt{conv$\doubleplus$relu 4}    & (13, 460, 230, 1, 1)  & (13, 345, 128, 5, 1)\\
\texttt{conv$\doubleplus$relu 5}    & (13, 230, 204, 7, 1)  & (13, 128, 307, 3, 1)  \\
\texttt{maxpool 3}                  & (13, 204, 3, 2)       & (13, 307, 3, 2)  \\
\texttt{global-pool 1}              & (1, 204)              & (1, 307) \\
\texttt{fc 1}                       & (204, 3686)           & (307, 3686) \\
\texttt{fc 2}                       & (3686, 6144)          & (3686, 6963) \\
\texttt{fc 3}                       & (3686, 1000)          & (3686, 1000) \\
\hline
Total energy (mJ)                   & 242.888               & 151.414 \\
\bottomrule
\end{tabular}
\label{tb:alextable}
\end{table}

\textbf{Applications-level.}
While the kernel- and model-level datasets can be beneficial for researchers and developers in understanding, modelling, and optimizing power and energy efficiency of DNN executions, end-users generally have a greater interest in the energy consumption of those frequently used AI applications on their devices. This is because the application's energy efficiency directly affects device's battery life, which is critical to the user experience.
To this end, we create an application-level dataset, which uncovers the end-to-end energy consumption of six popular edge AI applications, covering three main categories: vision-based (object detection, image classification, super resolution, and image segmentation), NLP-based (natural language question answering), and voice-based applications (speech recognition).
As shown in Table \ref{tab:appdata}, we measure the power and energy consumption of each application with multiple reference DNN models that operate under four distinct computational settings, including CPU with a single thread, CPU with four threads, GPU delegate, and the NNAPI delegate.
The dataset can serve as a resource for exploring the energy consumption distribution throughout the end-to-end processing pipeline of an edge AI application.
For example, we can use the dataset to examine the energy consumed in generating image frames, converting these frames from YUV to RGB, and conducting DNN inference within an object detection application. 
Fig. \ref{fig:e2e} depicts the energy consumption breakdown based on the processing phases in the object detection. 
It demonstrates that our application-level dataset can provide interpretable observations for comprehending who is the primary energy consumer in the end-to-end edge AI application. 
Additionally, the application-level dataset offers essential inputs for our edge device scoring system (Section \ref{sc:scoring}).
Due to the page limit, we will not present additional measurement results in this paper.

\begin{table*}
\centering
\caption{Measured edge AI applications per device in our application-level dataset.}
\vspace{-0.1in}
  \label{tab:appdata}
\resizebox{\textwidth}{!}
{
\begin{tabular}{ccccccccc}
\toprule
\multicolumn{1}{c}{}                                    & \multicolumn{1}{c}{}                                       & \multicolumn{1}{c}{}                               & \multicolumn{1}{c}{}                               & \multicolumn{4}{c}{Delegate}                                                                                            & \multicolumn{1}{c}{}                                                                                     \\ \cline{5-8}
\multicolumn{1}{c}{\multirow{-2}{*}{Category}} & \multicolumn{1}{c}{\multirow{-2}{*}{Application}} & \multicolumn{1}{c}{\multirow{-2}{*}{No.}} & \multicolumn{1}{c}{\multirow{-2}{*}{Reference DNN models}} & \multicolumn{1}{l}{CPU1} & \multicolumn{1}{l}{CPU4} & \multicolumn{1}{l}{GPU}                     & NNAPI                      & \multicolumn{1}{c}{\multirow{-2}{*}{\begin{tabular}[c]{@{}c@{}}Model size\\ (MB)\end{tabular}}} \\ \hline
                                                          &                                                             & DNN1                                                   & MobileNetv2, FP32, 300 × 300 pixels            & \multicolumn{1}{l} {\color{green} \checkmark}  & \multicolumn{1}{l} {\color{green} \checkmark}  & \multicolumn{1}{l}{}                        &  {\color{green} \checkmark}                         & 24.2                                                                                                      \\ \cline{3-9} 
                                                          &                                                             & DNN2                                                   & MobileNetv2, INT8, 300 × 300 pixels            & \multicolumn{1}{l} {\color{green} \checkmark}  & \multicolumn{1}{l} {\color{green} \checkmark}  & \multicolumn{1}{l}{}                        &  {\color{green} \checkmark}                         & 6.9                                                                                                       \\ \cline{3-9} 
                                                          &                                                             & DNN3                                                   & MobileNetv2, FP32, 640 × 640 pixels   & \multicolumn{1}{l} {\color{green} \checkmark}  & \multicolumn{1}{l} {\color{green} \checkmark}  & \multicolumn{1}{l}{}                        &  {\color{green} \checkmark}                         & 12.3                                                                                                      \\ \cline{3-9} 
                                                          & \multirow{-4}{*}{Image detection}                           & DNN4                                                   & MobileNetv2, INT8, 640 × 640 pixels   & \multicolumn{1}{l} {\color{green} \checkmark}  & \multicolumn{1}{l} {\color{green} \checkmark}  & \multicolumn{1}{l}{}                        &  {\color{green} \checkmark}                         & 4.5                                                                                                       \\ \cline{2-9} 
                                                          &                                                             & DNN5                                                   & EfficientNet, FP32, 224 × 224 pixels                & \multicolumn{1}{l} {\color{green} \checkmark}  & \multicolumn{1}{l} {\color{green} \checkmark}  & \multicolumn{1}{l} {\color{green} \checkmark}                     &  {\color{green} \checkmark}                         & 18.6                                                                                                      \\ \cline{3-9} 
                                                          &                                                             & DNN6                                                   & EfficientNet, INT8, 224 × 224 pixels                & \multicolumn{1}{l} {\color{green} \checkmark}  & \multicolumn{1}{l} {\color{green} \checkmark}  & \multicolumn{1}{l}{}                        &  {\color{green} \checkmark}                         & 5.4                                                                                                       \\ \cline{3-9} 
                                                          &                                                             & DNN7                                                   & MobileNetv1, FP32, 224 × 224 pixels                & \multicolumn{1}{l} {\color{green} \checkmark}  & \multicolumn{1}{l} {\color{green} \checkmark}  & \multicolumn{1}{l} {\color{green} \checkmark}                     &  {\color{green} \checkmark}                         & 4.3                                                                                                       \\ \cline{3-9} 
                                                          & \multirow{-4}{*}{Image classification}                      & DNN8                                                   & MobileNetv1, INT8, 224 × 224 pixels                & \multicolumn{1}{l} {\color{green} \checkmark}  & \multicolumn{1}{l} {\color{green} \checkmark}  & \multicolumn{1}{l}{}                        &  {\color{green} \checkmark}                         & 16.9                                                                                                      \\ \cline{2-9} 
                                                          & Super resolution                                            & DNN9                                                   & ESRGAN \cite{wang2018esrgan}, FP32, 50 × 50 pixels                         & \multicolumn{1}{l} {\color{green} \checkmark}  & \multicolumn{1}{l}{}     & \multicolumn{1}{l} {\color{green} \checkmark}                     &                            & 5                                                                                                         \\ \cline{2-9} 
\multirow{-10}{*}{Vision-based}            & Image segmentation                                          & DNN10                                                  & DeepLabv3 \cite{chen2018encoder}, FP32, 257 × 257 pixels                  & \multicolumn{1}{l}{}     & \multicolumn{1}{l} {\color{green} \checkmark}  & \multicolumn{1}{l}{}                        &                            & 2.8                                                                                                       \\ \hline
NLP-based                                  & Natural language question answering                                                   & DNN11                                                  & MobileBERT \cite{devlin2018bert}, FP32                                   & \multicolumn{1}{l} {\color{green} \checkmark}  & \multicolumn{1}{l} {\color{green} \checkmark}  & \multicolumn{1}{l}{}                        &  {\color{green} \checkmark}                         & 100.7                                                                                                     \\ \hline
Voice-based                                & Speech recognition                                          & DNN12                                                  & Conv-Actions-Frozen \cite{warden2018speech}, FP32                           & \multicolumn{1}{l} {\color{green} \checkmark}  & \multicolumn{1}{l} {\color{green} \checkmark}  & \multicolumn{1}{l}{{\color[HTML]{FE0000} }} & {\color{green}  {\color{green} \checkmark} } & 3.8                                                                                                       \\ \bottomrule
\end{tabular}
}
    
\end{table*}

\textbf{Time cost.}
Finally, in Table \ref{tb:time_cost}, we report the time cost associated with performing measurements and creating our datasets.
On a single edge device, we spend $23.1$, $4.7$, and $1.5$ days, respectively, on (1) measuring the power and energy consumption of all the generated kernels, DNN models, and edge AI applications, and (2) creating the corresponding power and energy datasets.
We will open-source our datasets and code for other researchers and developers. Collectively, we anticipate that the community will collaborate to create a larger scale energy dataset for a variety of edge devices.

\begin{table}[t]
\caption{Time cost of measurements per edge device.}
\vspace{-0.1in}
\small
\begin{tabular}{cccc}
\toprule
                            & Kernels       & Models     & Applications       \\ 
\hline 
\centering Measure time per device     &  $23.1$ days   &  $4.7$ days &     $1.5$ days         \\   
\bottomrule
\end{tabular}
\label{tb:time_cost}
\end{table}

\section{Energy Prediction}
\label{sc:pred}
In this section, we present our proposed solution to address \textbf{C2: extensibility}. 
To extend the applicability of our measurement study to a wider variety of DNN models, including those not present in our dataset, we design and implement a kernel-level energy predictor which can accurately predict the energy consumption of new DNN models on edge devices. The predictors are trained using our kernel-level dataset and evaluated by our model-level dataset.

\subsection{Design and Implementation}
Our designed kernel-level energy prediction method is inspired by nn-meter \cite{zhang2021nn} which proposed a kernel-based latency predictor for DNN models. However, nn-meter does not support energy prediction. We propose using a rationale akin to that of nn-meter for the design of our kernel-level energy predictor, especially given that kernels run sequentially on current edge devices. The key contributions of our proposed energy predictor include: (1) being the first energy predictor for modern edge devices, achieving an accuracy of 86.2\% (making it the most accurate energy predictor for edge devices to date) for unseen DNNs (i.e., those with unfamiliar kernel configurations); and (2) being the first kernel-level energy predictor for DNN executions on modern edge devices. Notably, most existing research primarily uses FLOPs to estimate the energy consumption of DNN executions, resulting in generally low prediction accuracy for unseen DNN models.
The core of our kernel-level energy prediction method is that we build and train a predictor for each type of kernel (e.g., conv$\doubleplus$bn$\doubleplus$relu) using the kernel-level energy dataset presented in Section \ref{sc:dataset}. The total energy consumption of a DNN model is then predicted by summing the estimated energy consumption of all kernels within that DNN model. 
Our energy predictors are implemented using the random forests regression, a machine learning algorithm known for its robustness, handling of high dimensional spaces, and its capability to model complex non-linear relationships.

\begin{figure*}[t]
\centering
\subfigure[Mobile CPU]
{\includegraphics[width=0.48\textwidth]{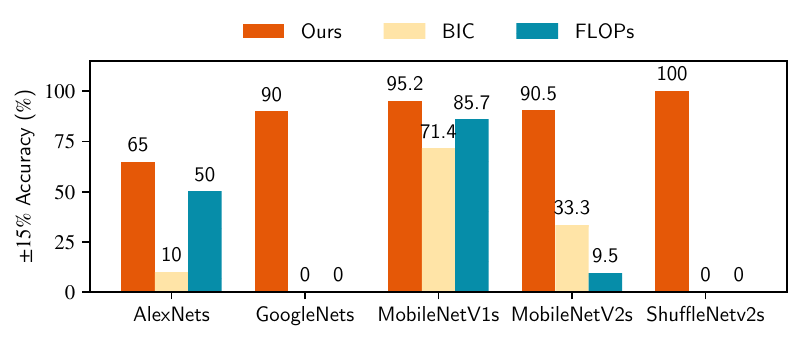}\label{fig:rescpu}}
\subfigure[Mobile GPU]
{\includegraphics[width=0.48\textwidth]{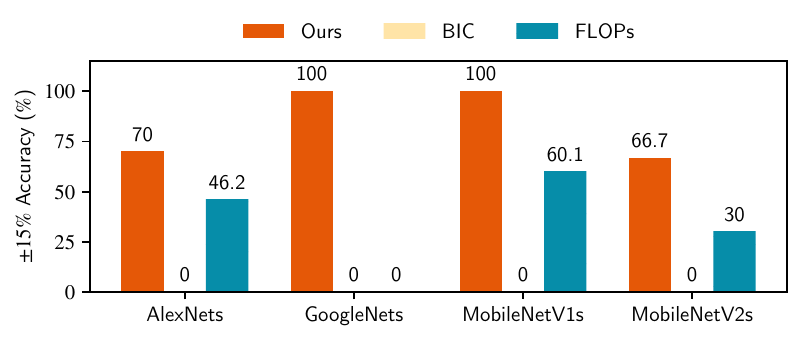}\label{fig:resgpu}}
\vspace{-0.1in}
\caption{Comparison of energy prediction performance. Our predictors trained by the kernel-level dataset achieves the highest accuracy on unseen DNNs.}
\label{fig:preres}   
\end{figure*}

\begin{table*}[t]
\centering
\caption{Energy prediction results on mobile CPU and GPU.}
\small
\begin{tabular}{ccccccccc}
\toprule
Model                               &\multicolumn{4}{c}{Mobile CPU}                &\multicolumn{4}{c}{Mobile GPU}\\
\cline{2-9}
variants                        &RMSE (mJ)  &RMSPE ($\%$)   &$\pm10\%$ (Acc.)  &$\pm15\%$ (Acc.) &RMSE (mJ)  &RMSPE ($\%$)   &$\pm10\%$ (Acc.)  &$\pm15\%$ (Acc.)\\\hline
AlexNets                        &32.2&12.9&60.0\%&65.0\%&9.4&15.5&40.0\%&70.0\%         \\
DenseNets                       &30.8&7.1&70.0\%&100\%  &16.5&19.6&10.0\%&35.0\%           \\
GoogleNets                      &25.1&11.9&20.0\%&90.0\%&3.8&5.5&95.0\%&100\%           \\
MobileNetv1s                    &7.8&8.7&80.9\%&95.2\%&1.7&6.7&80.9\%&100\%           \\
MobileNetv2s                    &7.8&8.3&76.2\%&90.5\%&3.1&11.5&47.6\%&66.7\%           \\
ProxylessNASs                   &13.3&11.7&47.6\%&71.4\%&2.5&8.2&76.2\%&95.2\%           \\
ResNet18s                       &44.6&6.1&95.2\%&100\%&30.5&13.1&38.1\%&71.0\%           \\
ShuffleNetv2s                   &3.2&5.8&100\%&100\%&-&-&-& -          \\
SqueezeNets                     &19.6&10.4&57.1\%&90.5\%&7.9&10.0&61.9\%&85.7\%           \\ \bottomrule
\end{tabular}
\label{tb:ours}
\end{table*}

\subsection{Performance Evaluation}
\textbf{Comparison baselines.} We implement two baselines to compare the energy prediction accuracy:
(1) FLOPs-based predictor: recent work has leveraged FLOPs to estimate the energy consumption of DNN inference \cite{desislavov2023trends}. We train FLOPs predictors using linear regression. Given the FLOPs of a DNN model, the predictor can estimate its inference energy consumption.
(2) BIC-based predictor: to demonstrate the critical role that our fine-grained kernel-level dataset plays in accurately predicting energy consumption, we also train energy predictors using the power data sampled by the edge device's built-in current (BIC) sensor.
To ensure a fair comparison (i.e, to confirm that any prediction errors in energy consumption are largely due to the inaccuracy of the built-in current sensor's power measurement), we take three steps: (1) using the ground-truth latency when calculating the energy consumption in the BIC training dataset, and (2) training the predictor with the same amount of data, covering the same number of kernels and identical configurations, and using the random forests regression.

\textbf{Metrics.}
The prediction performance is evaluated through the root mean square error (RMSE), root mean square percentage error (RMSPE), $\pm10\%$, and $\pm15\%$ accuracy. 
The latter two metrics represent the percentage of models whose predicted energy consumption lies within the specified error bounds relative to actual measured energy consumption.
In this paper, $\pm15\%$ accuracy is the default metric.
Smaller RMSE/RMSPE and larger $\pm10\%$/$\pm15\%$ indicate better prediction performance.

\textbf{Comparison results on unseen DNN models.} For the comparison study, we select AlexNets, GoogleNets, MobileNetv1s, MobileNetv2s, and ShuffleNetv2s. As the FLOPs-based predictor requires training with model-level data (i.e., the FLOPs of DNN models), we adopt a leave-one-out cross-validation approach. We set aside one model (e.g., 50 models of GoogleNets) as the test set, and use the remaining four models (e.g., 50 models each of AlexNets, MobileNetv1s, MobileNetv2s, and ShuffleNetv2s) as the training set to train the predictor. Our kernel-level predictor and the BIC-based predictor do not require model-level data for training. 

The comparison results are depicted in Fig. \ref{fig:preres}. 
Our kernel-level energy predictor consistently outperforms the other two baselines, delivering the highest prediction accuracy. Those baselines fail to achieve comparable levels of prediction accuracy on unseen DNN models.
Specifically, our predictor achieves an average prediction accuracy of $86.2\%$, significantly higher than FLOPs, $31.3\%$, and BIC, $12.7\%$.
The poor prediction accuracy of BIC, particularly on mobile GPU, demonstrates the indispensability of a fine-grained power and energy dataset when training a reliable energy predictor for edge devices.
The significant drop in prediction performance of BIC on the mobile GPU is due to the fact that DNNs typically achieve much shorter execution time on the GPU compared to the CPU. This shorter execution time on the GPU necessitates a higher power sampling rate.
Moreover, the performance gap between our kernel-level predictor and the FLOPs-based predictor reflects the gain derived through considering the runtime optimization of edge devices, such as kernel fusion. Table \ref{tb:ours} presents the prediction results evaluated across all nine DNN models in our model-level dataset.
In addition, we calculate the kernel configuration overlaps between the training (kernel-level dataset) and the evaluation (model-level detaset) datasets. Results show that our energy predictors have only seen $1.1\%$ (CPU) and $1.8\%$ (GPU) of the configurations in the evaluation dataset, which further attests the effectiveness of our kernel-level energy predictors on unseen models.

\textbf{Discussion.} Our kernel-level energy predictor exhibits slightly lower prediction accuracy compared to the latency predictor developed in nn-meter \cite{zhang2021nn}. This might primarily be due to the fact that (1) nn-meter manually sets CPU frequency of the measured device to a fixed value (2.42GHz) when profiling the latency for building the training dataset and evaluating the prediction accuracy. This creates a more controlled environment for latency measurement and prediction. However, to ensure practicality, our kernel-level energy predictor does not establish a fixed CPU frequency during energy measurement and prediction. This results in greater variability and potential uncertainty in the energy prediction, yet it more accurately reflects real-world usage scenarios where the CPU frequency is typically dynamic.
(2) The scale of our energy training dataset is less extensive than that of the latency training dataset in nn-meter,
as collecting fine-grained power data is significantly more time-consuming than profiling latency data, particularly on modern edge devices. Hence, we anticipate the community will collectively collaborate to further enhance the scale of our datasets.

\section{Scoring System}
\label{sc:scoring}
In this section, we introduce our method to tackle challenge \textbf{C3: understandability}. We develop a scoring system for diverse edge devices by leveraging our application-level dataset.
To ensure that the energy efficiency assessment result is accessible to a broad audience, in particular, edge device end-users with non-technical backgrounds, we develop two scoring metrics, namely \textit{power consumption score (PCS)} and \textit{inference energy consumption score (IECS)}. These two scoring metrics help to distill the power and energy efficiency of a device in an intuitive and understandable way.

\textbf{PCS.} The PCS is designed to capture the aggregated power efficiency (PE) for running all six edge AI applications with 12 reference DNN models using CPU, GPU, and NNAPI delegates. It is calculated as $PCS = \frac{\sum_{i=1}^n {PE}_i}{n}$, where $n$ is the total number of reference DNN models and $PE = (1 - \frac{APC}{TDP})\times 100$. APC denotes the average power consumption for inferences. Thermal design power (TDP), measured in watts, represents the maximum power an edge device is designed to consume under normal operating conditions. The ratio $\frac{APC}{TDP}$ provides an indication of how efficiently a device is using its power budget, with a lower ratio indicating better PE.

\textbf{IECS.} The IECS is designed to assess edge device energy efficiency, and calculated as the sum of inference energy consumption (IEC) for all six edge AI applications under CPU, GPU, and NNAPI delegates. IEC is defined as the number of inferences per unit of energy, where it factors in the trade-off between PE and inference latency. An edge device with a higher IECS is considered more energy-efficient. 

\begin{figure}[t]
\centerline{\includegraphics[width=0.48\textwidth]{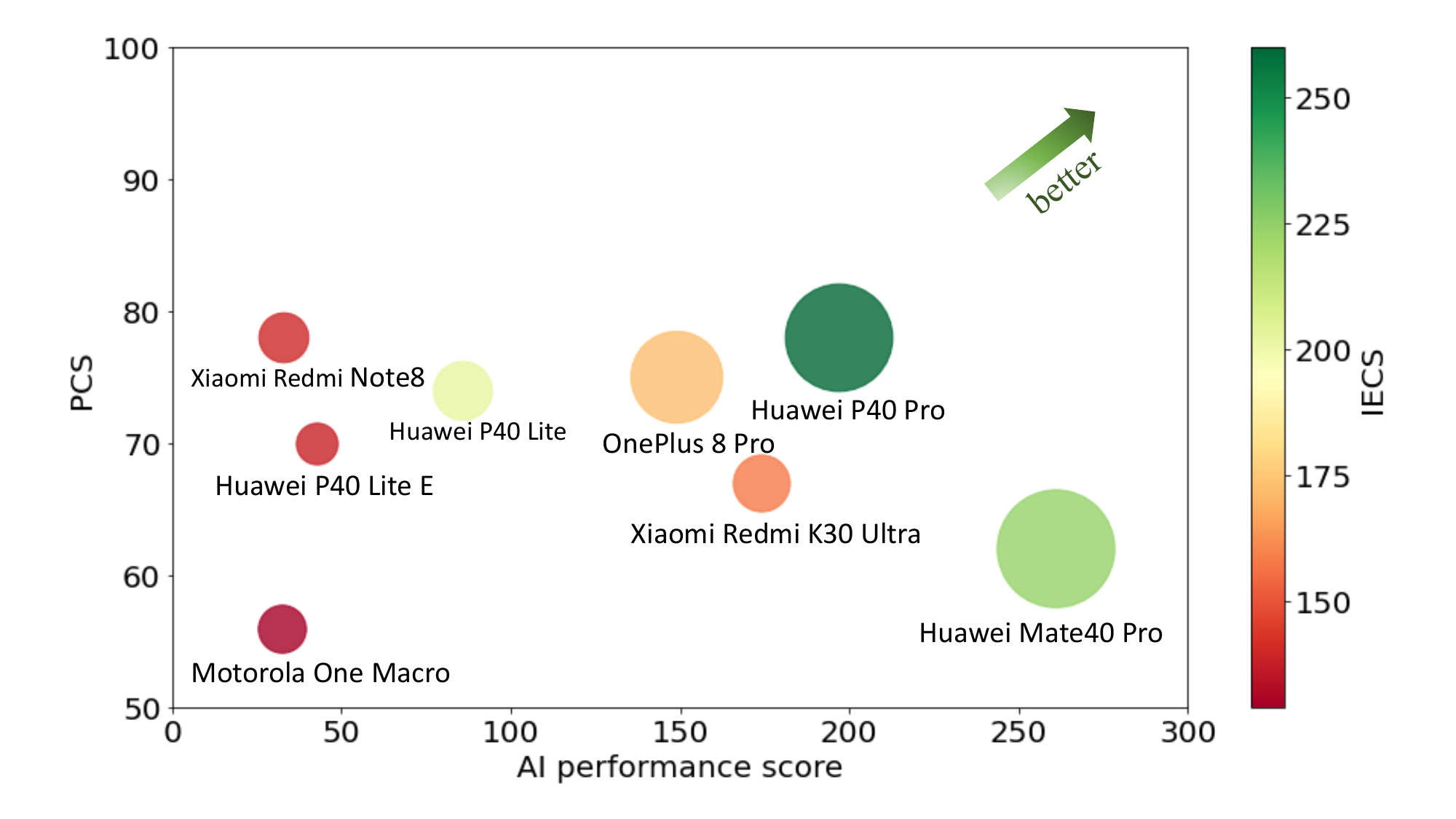}}
\vspace{-0.1in}
\caption{Comparison of the proposed PCS and AI inference score \cite{ignatov2018ai}. It presents a tradeoff among AI performance, power consumption, and the selling price. The larger the ball, the higher the selling price of the device.}
\label{fig:scatter}
\end{figure}

\textbf{Results.}
Fig. \ref{fig:scatter} compares our proposed PCS with the AI inference score developed by AI Benchmark \cite{ignatov2018ai} across diverse edge devices. Note that the AI inference score does not take into account power and energy efficiency.
The figure illustrates a tradeoff between AI performance, power consumption, and its selling price, where a larger ball in the figure represents a higher selling price for the device. An edge device that exhibits superior power efficiency (higher PCS) and AI inference performance (higher AI performance score) is positioned towards the top right corner of the figure.

We find that scoring metrics significantly influence benchmarking results for edge devices. For instance, although the Huawei Mate40 Pro achieves the highest AI performance score, it holds the second worst PCS. Conversely, the Xiaomi Redmi Note8 attains the highest PCS while having the second lowest AI performance score. These observations highlight the need for the development of IECS that balances power efficiency with AI inference performance. In Fig. \ref{fig:scatter}, the color of each ball indicates the IECS of each edge device. The Huawei P40 Pro presents the best equilibrium between AI performance and power efficiency, as indicated by its IECS and its position in the figure. \textit{The complete IECS benchmark results can be found on our project webpage}.

\section{Discussion}

\textbf{Limitations.} Our current measurements and datasets are on modern smartphones equipped with mobile CPUs and GPUs. While they cover a broad spectrum of edge hardware, they might not be comprehensive. To further increase the heterogeneity, we plan to extend our energy datasets by including other modern edge devices, such as Jetson Nano, Coral TPU, and Raspberry Pi 4.

The proposed kernel-level energy predictor is built offline and will not be
updated dynamically during DNN executions. Naturally, the prediction accuracy could be further improved by factoring in more environmental complexities, such as the available computing and memory resources on an edge device. We will leave this as an area for our future work.

\textbf{Automated measurement.}
Table \ref{tb:time_cost} illustrates that the majority of the time cost comes from energy profiling. Developing an automated measurement and profiling method can enhance the time efficiency for collecting a large-scale and more comprehensive dataset that includes a variety of edge devices and kernel configurations. The kernel-level energy predictor could also benefit, as prediction accuracy may improve with more training data. Furthermore, automated profiling could help minimize human influence, leading to more accurate measurements.

\textbf{Energy prediction for concurrent executions.}
Our energy predictor is premised on the fact that kernels currently run sequentially on edge devices. In the future, DNN inference may run concurrently on multi-core chipsets. Kernels processed in parallel might consume less energy than when processed sequentially, but more than individual kernels. The energy prediction performance for concurrent execution might be lower than for sequential execution, as concurrent operations introduce greater uncertainties in energy consumption. This aspect requires further experimentation.

\section{Related Work}
\textbf{Energy measurement for edge devices.}
A number of research works have proposed different methodologies and developed frameworks for measuring the energy consumption in mobile and edge devices. The Green Miner proposed in \cite{hindle2014greenminer} can physically measure the energy consumption of mobile devices such as Android phones and automate the testing of applications. The GfxDoctor developed in \cite{ding2017gfxdoctor} can systematically diagnose energy inefficiencies in app graphics at the app source-code level. However, none of these works have studied fine-grained energy measurement of DNNs on modern edge devices. 

\textbf{Edge AI benchmark.} A few recent studies developed mobile AI benchmarks that measure the performance of on-device training and inference. For example, AI Benchmark \cite{ignatov2018ai,ignatov2019ai} is arguably the first benchmark suite for mobile devices, which primarily focuses on Android smartphones and measures only the latency. MLPerf Mobile \cite{reddi2020mlperf, janapa2022mlperf} presents the first industry-standard open-source benchmark for performance and accuracy evaluation of mobile AI devices. Additionally, AIoTBench \cite{luo2020comparison} comprises a wider range of model architectures and AI frameworks, with a focus on assessing the inference capabilities of mobile and embedded devices. However, none of these edge AI benchmarks focused on energy efficiency of on-device learning and energy dataset creation for edge devices.

\section{Conclusion}
We conduct energy consumption measurement studies for on-device deep learning. We have created extensive energy datasets at the kernel-, model-, and application-level to facilitate research aimed at improving the energy efficiency of deep learning on edge devices. Building upon our energy datasets, we have developed kernel-level predictors that can accurately estimate the energy consumption of unseen DNN executions. Furthermore, we have implemented two scoring metrics to enhance the understandability of our energy measurement results. These contributions offer valuable resources for advancing energy-efficient deep learning on edge devices.

\begin{acks}
This work was supported by funds from Toyota Motor North America and by the US National Science Foundation (NSF) under Grant No. 1910667, 1910891, and 2025284.
\end{acks}

\bibliographystyle{unsrt}
\bibliography{references}

\end{document}